\documentclass[journal=jctcce,manuscript=article]{achemso}
\setkeys{acs}{articletitle=true}

\usepackage[version=3]{mhchem} 
\usepackage{mciteplus}
\usepackage{rotating}
\usepackage{amscd}
\usepackage{color}
\usepackage{subcaption}
\usepackage[usenames,dvipsnames]{xcolor}
\usepackage{setspace}
\usepackage{xspace}
\usepackage{xr}
\usepackage{hyperref}

\newcommand*{\kcal}{kcal\,mol$^{-1}$\xspace}

\newcommand*{\degrees}{$^{\circ}$\xspace}
\newcommand*{\invcm}{cm$^{-1}$\xspace}

\author{Kavindri Ranasinghe}
\affiliation{Recursion, Schr\"odinger Building, Oxford Science Park, Oxford, UK}
\alsoaffiliation{Exscientia, Schr\"odinger Building, Oxford Science Park, Oxford, UK}
\altaffiliation{Contributed equally}
\author{Adam L. Baskerville}
\affiliation{Exscientia, Schr\"odinger Building, Oxford Science Park, Oxford, UK}
\altaffiliation{Contributed equally}
\author{Geoffrey P. F. Wood}
\affiliation{Recursion, Schr\"odinger Building, Oxford Science Park, Oxford, UK}
\alsoaffiliation{Exscientia, Schr\"odinger Building, Oxford Science Park, Oxford, UK}
\author{Gerhard K\"onig}
\affiliation{Recursion, Schr\"odinger Building, Oxford Science Park, Oxford, UK}
\alsoaffiliation{Exscientia, Schr\"odinger Building, Oxford Science Park, Oxford, UK}
\email{gerhard.koenig@recursion.com}

\title[]{Basic stability tests of machine learning potentials for molecular simulations in computational drug discovery}

\begin{document}
\maketitle


\begin{abstract}
Neural network potentials trained on quantum-mechanical data can calculate molecular interactions with relatively high speed and accuracy. However, not all neural network potentials are suitable for molecular simulations, as they might exhibit instabilities, nonphysical behavior, or lack accuracy. To assess the reliability of neural network potentials, a series of tests is conducted during model training, in the gas phase, and in the condensed phase. The testing procedure is performed for eight in-house neural network potentials based on the ANI-2x dataset, using both the ANI-2x and MACE architectures. This consistent framework allows an evaluation of the effect of the model architecture on its performance. For comparison, we also perform stability tests of the publicly available neural network potentials: ANI-2x, ANI-1ccx, MACE-OFF23, and AIMNet2. The results show that the different models have different weaknesses. A normal mode analysis of 14 simple benchmark molecules with large displacements from the energy minima revealed that the published small MACE-OFF23 model shows large deviations from the reference quantum-mechanical energy surface. Also, some MACE models with a reduced number of parameters failed to produce stable molecular dynamics simulations in the gas phase, and all MACE models exhibit unfavorable behavior during steric clashes. In addition, the published ANI-2x and one of the in-house MACE models are not able to reproduce the structure of liquid water at ambient conditions, forming an amorphous solid phase instead. For the ANI-1ccx model, the multi-body interactions in the condensed water phase lead to nonphysical additional energy minima in bond length and bond angle space, which caused a phase transition to an amorphous solid. Out of all 13 considered public and in-house models, only one in-house model based on the ANI-2x B97-3c dataset shows better agreement with the experimental radial distribution function of water than the simple molecular mechanics TIP3P model. The mixed results for the machine learning potentials show that great care must be taken during model training and when selecting a neural network potential for real-world applications.  

\end{abstract}

\section{Introduction}
\label{sec:introduction}

The use of machine learning (ML) has emerged as a promising approach for molecular simulations. In particular neural network potentials (NNPs) that were trained on quantum-mechanical data might offer quantum-chemical accuracy with computational costs that are comparable to classical molecular mechanics force fields. This also makes them attractive for applications in computational drug discovery, like predictions of protein-ligand binding, as well as other important drug properties like solubility, chemical and physical stability and permeability. However, before NNPs can be included in automated drug discovery pipelines, their strengths and weaknesses need to be assessed to ensure reliable results.    

The quality of NNPs depends on a series of factors during model development like the underlying dataset of quantum-chemical calculations, the neural network architecture, the selected features, the data split into training, validation and test sets, the employment of energies and forces for training among others. Initially, the performance of NNPs is measured using an error metric with respect to the testing data. However, most of the model testing is performed in the gas-phase using a limited set of molecules and conformations; therefore, it is unclear how far the models are transferable to other application areas such as simulations in the condensed phase, commonplace in computational drug discovery.    

Significant advances in NNP development have been achieved in recent years, including ANI,\cite{ani1,ani1ccx,ani2x} MACE,\cite{mace,maceoff23} or AIMNet2\cite{aimnet2}  emerging as prominent examples. This also raises the question of which model and which architecture is most suitable for applications in computational drug discovery? Comparing different architectures is particularly challenging, as most models are trained on distinct datasets. Consequently, it is unclear whether observed differences can be attributed to the training set or the model architecture.    

\begin{table}[ht!]
    \caption{Overview of the created in-house models based on the ANI-2x and MACE architectures,\cite{ani2x,mace} and trained on the published ANI-2x datasets.\cite{ani2x_data}   } 
    \label{tab:models}
    \centering    
    \begin{tabular}{|l|l|l|}
    \hline 
    Model name        & Model architecture & Training set \\ 
    \hline 
        ANI-2x RXRX 1  &  ANI-2x & ANI-2x $\omega$B97X/6-31G(d) \\
        ANI-2x RXRX 2  &  ANI-2x & ANI-2x $\omega$B97X/6-31G(d) \\
        ANI-2x RXRX 3  &  ANI-2x & ANI-2x $\omega$B97X/6-31G(d) \\
        ANI-2x RXRX 4  &  ANI-2x & ANI-2x $\omega$B97X/6-31G(d) \\
        \hline
        MACE RXRX XXS  &  MACE   & ANI-2x $\omega$B97X/6-31G(d) \\
        MACE RXRX XXXS &  MACE   & ANI-2x $\omega$B97X/6-31G(d) \\ 
        \hline 
        B97-3c RXRX 1  &  ANI-2x & ANI-2x B97-3c/def2-mTZVP \\
        B97-3c RXRX 2  &  ANI-2x & ANI-2x B97-3c/def2-mTZVP \\
        \hline 
    \end{tabular}
\end{table}

To evaluate the stability of different models and architectures, we employ both publicly available models and models generated in-house that use a consistent data set and data split during model training. Our in-house models enable for a fair comparison of different model architectures and the error metrics from their training stages can be compared with other error metrics. This allows an early identification of possible issues at the training stage. In total, eight different in-house models were generated (see Table~\ref{tab:models}). Six models were trained on the ANI-2x $\omega$B97X/6-31G(d) dataset,\cite{ani2x_data} and used either the ANI-2x or MACE architectures.\cite{ani2x,mace} To test the uncertainty of model training, four models with the ANI-2x architecture were generated that differ only in the random seeds during training (ANI-2x RXRX 1 to 4). Because of the high computational costs, the two models based on the MACE architecture are much smaller than the published MACE models and use different numbers of parameters (MACE RXRX XXS and XXXS, see Table~\ref{tab:EXX-MACE-models}). To illustrate the effect of using a training set with a  different quantum-mechanical level of theory, two models with the ANI-2x architecture were trained on the ANI-2x B97-3c/def2-mTZVP  dataset (B97-3c RXRX 1 and 2).\cite{ani2x_data} As a reference to the in-house models we used ANI-2x,\cite{ani2x} ANI-1ccx,\cite{ani1ccx} AIMNet2,\cite{aimnet2} and three MACE-OFF23 models with size S, M, and L,\cite{maceoff23} as implemented in torchani.\cite{torchani} 

In the following, we present a series of basic tests to evaluate the performance of NNPs in areas that they were not trained for. The tests attempt to answer the question to what extent the models describe the fundamental physics, and whether some models perform consistently better across a wide range of applications. This is done in different stages. First, we employ the in-house models to check some error metrics of the models during the training phase. Then we progress to benchmarks in the gas phase, focusing on normal mode analyses of 14 simple molecules, molecular dynamics simulations of a large drug-like benchmark molecule, as well as hydrogen bonding and van der Waals interactions of water dimers. The final test is performed in the condensed phase, looking at the densities and radial distribution functions of the models for water. This is motivated by the importance of water-water interactions for the hydrophobic effect, which drives protein-ligand binding.    

\section{Methods}
\label{sec:methods}

\subsection{Model training}
\texttt{PhysicsML} (\url{https://github.com/Exscientia/physicsml}) is a package developed by researchers at Exscientia for handling physics-based NNPs. Building upon \texttt{molflux} (\url{https://github.com/Exscientia/molflux}), a foundational package for molecular predictive modeling, \texttt{PhysicsML} offers a standardized interface for building and training machine learning models from scratch. The in-house models introduced in this work were developed using the \texttt{PhysicsML} package, incorporating Data Version Control (DVC) (\url{https://github.com/iterative/dvc}) for version control. More information on training a \texttt{PhysicsML} model can be found in the following tutorial: \url{https://exscientia.github.io/physicsml/pages/tutorials/ani1x_energy_forces_training.html}.

The ANI-2x dataset, stored in the \texttt{molflux.dataset} catalog, was initially loaded and then split using a shuffle split (random permutation cross-validation) method into 75\% for training, 5\% for validation, and 20\% for testing. This split ensures that the data is randomly divided, offering a fair representation in each subset, and the same split was applied to all models. Given the \texttt{mol\_bytes} format, an RDKit serialization of the 3D molecules, we then featurized the dataset using the \texttt{PhysicsML} featurizer and extracted the atomic numbers, coordinates, and self-atomic energies (SAEs). During model training, the \texttt{PhysicsML} models are accessed via \texttt{molflux.modelzoo} module, which handles the basic standard API. The models were trained on both energies and forces, with units of \kcal{} and \kcal{\AA}$^{-1}$ , respectively. The model training was conducted on Amazon EC2 g5-12xlarge instances, using four GPUs per model. All models were pre-batched to avoid any data loading effects, with batch sizes chosen to fully utilize GPU memory.

To ensure consistency, we utilized the same model architectures as the published models of ANI-2x and MACE-OFF23\cite{ani2x,maceoff23}. However, an extensive hyperparameter tuning exercise was initially performed on selected parameters to identify the optimal settings for achieving maximum accuracy and computationally efficient model training.
In this analysis, we implemented a convergence criterion for halting the model training, preventing overfitting and unnecessary computation. The training process was monitored based on the performance metrics, validation loss and accuracy. We utilized the \texttt{ReduceLROnPlateau} callback to dynamically adjust the learning rate based on the model's performance. If the validation loss did not improve for 50 epochs (as defined by the patience parameter), the learning rate was reduced by a factor of 0.8 to help the model continue learning effectively. The \texttt{LearningRateMonitor} in the PyTorch Lightning framework was used with a \texttt{min\_delta} of 0.01, setting a threshold for the minimum change in the monitored metric to qualify as an improvement. If the change in validation loss or accuracy was less than this threshold, it was not considered significant. Additionally, an \texttt{EarlyStopping} callback from PyTorch Lightning was employed to monitor the validation loss during model training and halt the process when the model's performance plateaued.

In atomistic neural network potentials, self-atomic energies (SAEs) are crucial as they represent the energy contribution of individual atoms within a molecule. By accurately calculating these energies, the model ensures better precision and numerical stability in the predicted total molecular energies. In this work, SAEs were computed using a least-squares regression methodology. We also applied data scaling parameters to standardize the range of data features, using the mean and standard deviation of the dataset.

During model training, we utilized the gradient clipping technique to prevent exploding gradients. This method helps in maintaining numerical stability and ensuring effective learning by capping the gradients at a maximum threshold, preventing them from growing too large. The choice between 32-bit and 64-bit precision can significantly impact both performance and accuracy of a model. In this study, all in-house models were trained using 32-bit single precision. This approach uses less memory, enabling larger batch sizes to be trained on the same hardware, faster computation and better utilization of GPU resources.

The AdamW optimization algorithm, which relies on moving averages of the gradients and their squares in its denominator, can run into issues when gradients become very small. When this happens, the already small denominator can lead to disproportionately large update steps. As a result, the model parameters can be pushed far from their optimal values, causing a significant increase in loss. This was observed when training with a very large unit of energy such as Hartrees (1 Hartree $\approx$ 627.509\,\kcal{}). Switching to \kcal{} provided a more manageable scale for numerical computations, reduced the risk of numerical instability, and improved the precision of calculations.

The MACE-OFF23\cite{maceoff23} models, built on the MACE\cite{mace} architecture has been previously shown to accurately predict gas and condensed phase properties of molecular systems, including dihedral torsion scans and descriptions of molecular crystals and liquids. Three models were reported: small, medium and large (abbreviated as S, M, and L), which indicate the number of hyperparameters used. In this work the number of hyperparameters is reduced even further in two smaller models, RXRX XXS and RXRX XXXS, where Table \ref{tab:EXX-MACE-models} reports the number of hyperparameters used in each model compared alongside those used in the MACE-OFF23 models. More information on the hyperparameter search is provided in the Supporting Information (SI).

\begin{table}[ht!]
    \caption{Key model parameters the in-house models, RXRX XXXS, RXRX XXS, which were trained using the original MACE architecture with differing hyperparameter initialisation. The hyperparameters for the MACE-OFF23 small, medium and large models\cite{maceoff23} are listed for comparison.}
    \label{tab:EXX-MACE-models}
    \centering
    \begin{tabular}{ccccccc}
                            &   RXRX XXXS & RXRX XXS & Small & Medium & Large \\ \hline
    Cutoff radius (\AA)     &   5.0      & 5.0       &  4.5                  &  5.0                   & 5.0   \\
    Chemical channels, $k$  &   16.      & 32.        &  96.                  &  128.                  & 192.  \\
    Correlation     &   3.          &   3.              &  3.                   &  3.                    & 3.    \\
    max $L$     &   0.          &   0.             &  0.                   &  1.                    & 2.    \\ \hline
    \end{tabular}

\end{table}


\subsection{Gas phase calculations}

\subsubsection{Normal mode analysis}

To understand the behavior and sensitivity of NNPs when they are presented structures away from equilibrium, we have performed an extensive normal mode analysis for the 14 small molecules give in Figure~\ref{fig:gastest}.  First, a single conformer of each molecule was generated via RDKIT.\cite{rdkit} For flexible molecules, no attempt was made to find the global minimum conformation but the conformer generation seed for the ETKDG version 3 algorithm\cite{wang2020improving} was set to 
\textit{0xF00D} to ensure reproducibility of the results. From this initial conformer the ORCA6\cite{orca} program package was used to optimize the structure into the nearest local minimum using the same level of theory used to train each NNP (see Table \ref{tab:gastest}). This was followed by a frequency calculation with the same level of theory used for the geometry optimizations.  The generated frequencies were checked for no imaginary frequencies to ensure that the geometry was in a true local minium.  After the frequency calculation the normal mode eigenvectors were used to construct a series of structures to follow each mode in both directions, forward and reverse, from the energy minimum. This was done by applying 20 sequential steps along the normal mode eigenvectors using step sizes of  $\pm0.1$ for modes with fundamental frequencies $\omega$ $<$ $500$\invcm{},  $\pm0.08$ for  $500$ $\le$ $\omega$ $<$ $1000$\,\invcm{}, $\pm0.04$ $1000$ $\le$ $\omega$ $<$  $1800$\,\invcm{}, and $\pm0.015$  $\omega$ $\ge$ $1800$\,\invcm{}.  With these structures generated, ORCA was used to perform single point energy evaluations at the given level of theory.  For each evaluated NNP an energy evaluation using the ASE infrastructure was performed on each structure. An analysis of these energy evaluations against the benchmark data is given in Table \ref{tab:normalmode}.

\begin{figure}[t!]
\caption{14 benchmark molecules for the normal mode analysis tests in the gas phase: Water, ethane, methanol, methanethiol, ethanol, acetamide, tetrahydrofuran, n-hexane, cyclohexane, benzene, phenol, aniline, N-acetyl-alanine-methylamide, and N-acetyl-serine-methylamide. 
 }\label{fig:gastest}
   \begin{tabular}{c}
      \includegraphics[width=0.5\linewidth]{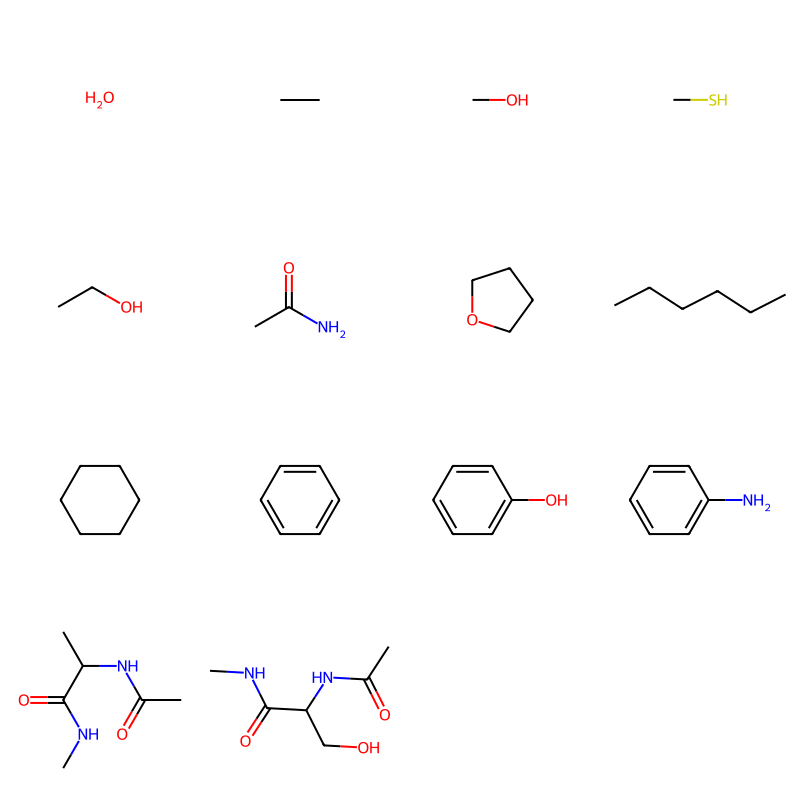}    
   \end{tabular}
\end{figure}

\subsubsection{Molecular Dynamics Simulations}

\begin{figure}[ht!]
\caption{Drug-like benchmark molecule for the MD simulation tests in the gas phase. The artificial molecule is a mixture of elements of clarithromycin, dexamethasone, diazepam, morphine, penicillin, sildenafil, and tryptophan dipeptide. The tests involves the stability of the molecule with $349$ atoms at $400$\,K, as well as the stretching of $370$ chemical bonds and the bending of $700$ bond angles.  
 }\label{fig:gastest}
   \begin{tabular}{c}
      \includegraphics[width=0.75\linewidth]{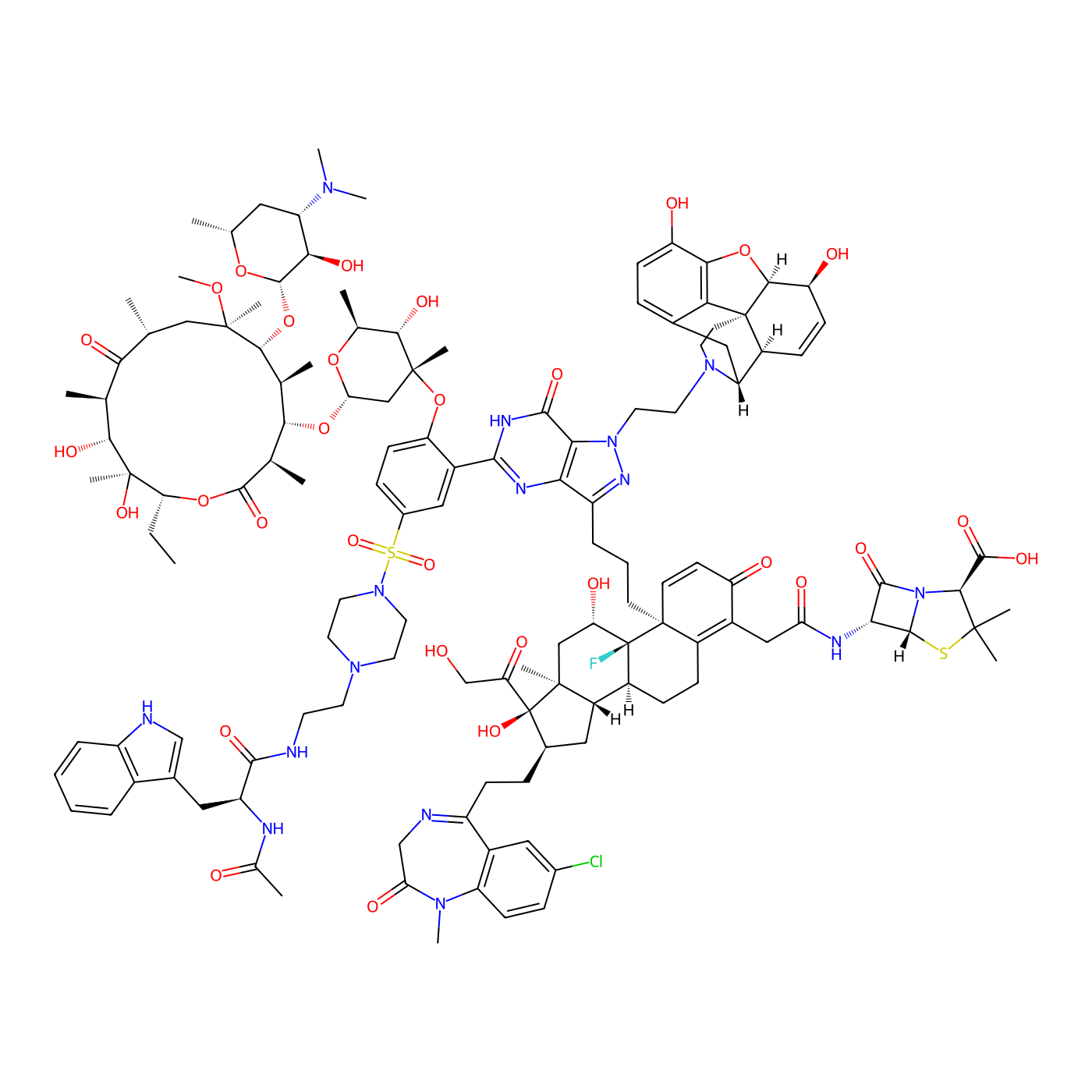}    
   \end{tabular}
\end{figure}

To represent some important parts of chemical space for computational drug discovery, an artificial benchmark molecule was created that contains structural features of clarithromycin, dexamethasone, diazepam, morphine, penicillin, sildenafil, as well as tryptophan dipeptide (Fig.~\ref{fig:gastest}). The SI also contains $14$ simple benchmark systems that were selected from previous publications and represent relevant chemical groups and levels of hydrophobicity and aromaticity.\cite{mybutanol,qmdrude,leds2} The benchmark molecule was generated based on the SMILE strings of the parent molecules using the GAFF2 force field.\cite{gaff2} Before each ML simulation, the geometry of the molecule was optimized using the L-BFGS minimizer. The molecular dynamics simulations at the ML level were performed with OpenMM 8.1\cite{Eastman2023a} at an elevated temperature of $400$\,K to test the stability of the molecules. Langevin dynamics were performed with a friction constant of $1$\,ps$^{-1}$.\cite{Chow1995} The time step was $1.0$\,fs and the simulation length $0.25$\,ns, except for the MACE M and L models, which only reached simulation lengths of $0.174$ and $0.041$\,ns within $24$\,h. The trajectory was saved every $0.5$\,ps. The distributions of all bond length and bond angle values during the simulation were calculated using MDTraj.\cite{mdtraj} The quasi-harmonic analysis of the individual distributions was performed as outlined in Ref.~\citenum{bro95a}. 

\subsubsection{Water dimer}

Starting from the Smith stationary point 1,\cite{gillan2016perspective} an energy optimisation was performed with $\omega$B97X/6-31G(d) using ORCA.\cite{orca5} For the reference results, a scan of the O-O distance was performed between $0.02$ and $0.70$\,nm with $\omega$B97X/6-31G(d). To ensure convergence, the maximum number of iterations for the SCF cycle and the optimization was increased to 10,000. An additional scan was performed with B97-3c/def2-mTZVP. The distance scans between $0.02$ and $0.4$\,nm with the ML potentials were performed with OpenMM 8.1,\cite{Eastman2023a}  using distance restraints with a strong force constant of 10\,GJ mol$^{-1}$ nm$^{-2}$. A geometry optimization was performed every $0.01$\,nm using the L-BFGS minimizer.  

\subsection{Condensed phase calculations}

The molecular dynamics simulations were performed with OpenMM 8.1.\cite{Eastman2023a}  The water box was first prepared with molecular mechanics, using 168 TIP3P water molecules.\cite{Jorgensen1983} The box was equilibrated for 0.5\,ns, leading to a box size of 1.706\,nm. The electrostatic interactions were computed with the particle mesh Ewald method,\cite{pme,Essmann1995} using a short-range cutoff of 8\,{\AA}.  The chemical bonds of TIP3P water were constrained with the SETTLE algorithm.\cite{Miyamoto1992} Before each ML simulation, the geometry of the solute was optimized using the L-BFGS minimizer, followed by $125$\,ps of constant volume equilibration. The  production simulation with constant pressure was performed for $0.125$\,ns, using a saving frequency of $0.5$\,ps for the trajectory. 
All simulations were conducted at a temperature of $300$\,K, using a Nose-Hoover thermostat,\cite{ zhang2019unified} and a pressure of $1$\,bar, using a Monte Carlo barostat.\cite{Aaqvist2004, Chow1995}  The time step was $0.5$\,fs. The radial distribution functions were calculated using MDTraj.\cite{mdtraj}

\section{Results and Discussion}
\label{sec:results}

\subsection{Model training stage} 

\begin{table}[ht!]
    \caption{Overview of the total number of epochs that were used for model training for each in-house model, as well as the corresponding final root mean square error (RMSE) with respect to the test data set.} 
    \label{tab:rmse}
    \centering
    \begin{tabular}{|lcc|}
    \hline 
    Model name        & Epochs & RMSE [\kcal{}] \\ 
\hline
\multicolumn{3}{|c|}{Trained on ANI-2x dataset} \\
\hline 
        ANI-2x RXRX 1  & 4957  & 1.8 \\
        ANI-2x RXRX 2  & 2458  & 1.9 \\
        ANI-2x RXRX 3  & 2430  & 1.9 \\
        ANI-2x RXRX 4  & 2199  & 1.9 \\
        \hline
        MACE RXRX XXS  &  3550 & 1.8 \\
        MACE RXRX XXXS &  5075 & 2.5 \\ 
\hline
\multicolumn{3}{|c|}{Trained on other datasets} \\
\hline 
        B97-3c RXRX 1  & 2575  & 1.8 \\
        B97-3c RXRX 2  & 1500  & 1.8 \\
        \hline 
    \end{tabular}
\end{table}

The number of epochs of model training until convergence, as well as the root mean square errors (RMSE) with respect to the QM single point energies of the test data set are listed in Table~\ref{tab:rmse}. The corresponding plots of the loss functions can be found in the Supporting Information (Figures S1 to S7). Overall, most models converged within about 2500 steps to an RMSE of about 1.9\,\kcal{}. This RMSE is comparable to the reported RMSE for the published ANI-2x model (1.9\,\kcal{} for relative conformer energies, and $1.8$\,\kcal{} for interaction energies based on the X40 data set).\cite{ani2x} There are only four models with a very different number of epochs for model training: (a) ANI-2x RXRX 1, which required 4957 epochs to reach the convergence criteria because the learning rate was raised to 0.001 after 1500 epochs to test the effect of a tempering strategy on model performance (see Figure~S1 in the Supporting Information). The resulting RMSE of ANI-2x RXRX 1 is slightly lower than the RMSE of ANI-2x RXRX 2 to 4 (1.8 versus 1.9 \kcal{}), but the computational costs of model training are twice as high. In the following tests, no differences were observed between ANI-2x RXRX 1 and ANI-2x RXRX 2 to 4, which indicates that the tempering step did not significantly improve model performance. (b) The second exception is model B97-3c RXRX 2, which was stopped after a fixed time of 1500 epochs. This model was intended to check the effect of using a fixed number of epochs rather than a convergence criterion based on the validation total loss. As can be seen in Table~\ref{tab:rmse}, both B97-3c RXRX 1 and B97-3c RXRX 2 yield the same RMSD (1.8\,\kcal{}), which indicates that a smaller number of epochs would have been sufficient. 

The two other models that required more epochs to reach convergence use the MACE architecture. (c) The MACE RXRX XXS  model required 3550 epochs to reach an RMSE of  1.8\,\kcal{}, thus being one of the four models with the lowest RMSE.  (d)  The MACE RXRX XXXS  model required 5075 epochs to reach an RMSE of 2.5\,\kcal{}, which is the largest deviation observed among all models. Like for ANI-2x RXRX 1, the training of this model involved a tempering step with an increased learning rate after 1200 epochs, which almost doubled the necessary number of epochs for convergence. Because of the high number of employed epochs, the poor model performance can probably be attributed to the lower number of parameters that are employed in this model (see Table~\ref{tab:EXX-MACE-models}). With the exception of MACE RXRX XXXS, all in-house models perform very similarly in terms of RMSE (between 1.8 and 1.9\,\kcal{}). Thus, no clear advantage of the MACE architecture compared to the ANI architecture was observed for the test set.  

\subsection{Gas phase}

\subsubsection{Normal mode analysis}  

One of the key aspects for molecular dynamics to perform well is the ability to faithfully simulate molecules away from their local and global minima structures, making sure that the energy surface is smooth and continuous for force-evaluations. To systematically test structures with increasing distances from the energy minimum, a normal mode analysis was performed for the 14 benchmark molecules in Fig.~\ref{fig:gastest}. Each normal mode was followed by stepping out from the localized minimum structure using the eigenvector matrix found for each mode.  The mode was followed in the forward and reverse directions by adding or subtracting an integer multiple of the matrix multiplied by the step size to the localized coordinates. The step size was selected so that low frequency modes climbed in energy of roughly the same amount as the high frequency modes. An initial analysis confirmed that no systematic differences of the NNPs exists across frequency ranges (see SI Tables S2 to S5), which indicates that their performance can be evaluated based on the average error over all frequencies. 

Table~\ref{tab:normalmode} shows the RMSE of the normal mode analysis for the different NNPs, using as the reference the same level of theory that was used for training. The RMSE was determined for all normal mode profiles against the benchmark data when reaching a certain relative energy cutoff of 50, 100, 200 and 400\,\kcal{} (denoted as CUT50, CUT100, CUT200, and CUT400, respectively). The RMSE progressively gets larger for all tested NNPs the higher the relative energy becomes. For example, for the published ANI-2x model in the first row, the RMSE for normal mode profiles below a cutoff of 50\,\kcal{} is 0.5\,\kcal{}, whereas for below a cutoff of 400\,\kcal{} the RMSE is 3.3 \kcal{}.  With CUT50, every NNP exhibits an RMSE of less than 1.2 \kcal. For the higher cutoff value, CUT400, the performance for most of the NNPs remains robust with RMSE values of less than 5.6\kcal. For the intermediate cutoffs (CUT100 and CUT200), the performance is scaled between the lower and higher RMSE values in each case. Interestingly, the NNPs trained on the ANI-2x data set tend to exhibit lower RMSE values than models that were trained on other datasets. This highlights that  dataset selection is as important as the architecture that was used. The data set size does not seem to be factor. For example, SPICEv2 was used for "MACE OFF M" which contains about 2 million data points, whereas, for AIMNET2, the dataset size is 20.3 million, and the ANI-2x data set size is roughly in between these sizes with 8 million. 

The notable exception in performance is for the MACE OFF S model for CUT400. In this case the RMSE is 75.6 \kcal{}, which is approximately \(13 \times\) higher then the next worst performing NNP.  This indicates that, at a certain cutoff value, some NNPs might break down and result in severe errors. For the "MACE OFF S" this happens at approximately 400 \kcal for the other models this occurs at a relative energy of greater than 400 \kcal{}, see SI. 

\begin{table}[!ht]
    \centering
    \caption{RMSE in \kcal{} from normal mode analysis of the 14 benchmark molecules, using four different cutoff values. The columns list the model, the level of theory for the reference calculations, the RMSE for all wave numbers using an energy cutoff for the displacements of 50\,\kcal{} (CUT50), 100\,\kcal{} (CUT100), 200\,\kcal{} (CUT200), and 400\,\kcal{} (CUT400). Larger energy cutoff values test larger displacements from the energy minimum along each normal mode.}\label{tab:normalmode}
        \begin{tabular}{|llcccc|}
    \hline
        Model & Reference & CUT50 &	CUT100	& CUT200 & CUT400  \\	
\hline
\multicolumn{6}{|c|}{Trained on ANI-2x dataset} \\
\hline 
ANI-2x	        &	$\omega$B97X/6-31G(d)	&	0.5	&	1.0	&	1.9	&	3.3\\
ANI-2x RXRX 1	&	$\omega$B97X/6-31G(d)	&	0.5	&	1.1	&	2.0 &	2.8\\
ANI-2x RXRX 2	&	$\omega$B97X/6-31G(d)	&	0.6	&	1.2	&	2.0 &	3.2	\\
ANI-2x RXRX 3	&	$\omega$B97X/6-31G(d)	&	0.5	&	1.1	&	1.9 &	3.3	\\
ANI-2x RXRX 4	&	$\omega$B97X/6-31G(d)	&	0.7	&	1.4	&	2.1 &	3.1	\\
\hline   
MACE RXRX XXS	&	$\omega$B97X/6-31G(d)	&	0.7	&	1.2	&	1.9 &	2.4	\\
MACE RXRX XXXS	&	$\omega$B97X/6-31G(d)	&	0.4	&	0.9	&	1.5 &	2.0	\\
\hline     
\multicolumn{6}{|c|}{Trained on other datasets} \\
\hline
AIMNET2	        &	B97-3c&	0.9	&	1.3	&	1.8	&	2.6 \\
B97-3c RXRX 1	&	B97-3c&	0.7	&	1.3	&	2.1	&	3.7 \\
B97-3c RXRX 2	&	B97-3c&	0.7	&	1.2	&	2.0 &	3.0	\\
MACE OFF S	    &	$\omega$B97M-D3BJ/def2-TZVPPD&	1.2	&	2.6	&	4.3	&	75.6 \\
MACE OFF M	    &	$\omega$B97M-D3BJ/def2-TZVPPD&	1.0	&	2.0	&	4.2	&	5.6 \\
\hline
    \end{tabular}
\end{table}

\subsubsection{Molecular dynamics simulations}  

\begin{table}[ht!]
\caption{Analysis of molecular dynamics simulations of the benchmark molecule from Fig.~\ref{fig:gastest} in the gas phase. Models that lead to unstable molecular dynamics simulations are marked with X. The average relative deviations ($\langle \Delta \rangle$) over all chemical bonds and angles with respect to the GAFF2 force field parameters are given in percent. Both the average values of the bond length ($r_{eq}$) and bond angle ($\theta_{eq}$)  are listed.}\label{tab:gastest}
\begin{tabular}{|l|l|l|}
\hline
\textbf{Model}     & \textbf{$\mathbf{\langle \Delta r_{eq} \rangle}$}  & \textbf{$\mathbf{\langle \Delta \theta_{eq} \rangle}$}   \\
\hline  
\multicolumn{3}{|c|}{Trained on ANI-2x dataset} \\ 
\hline 
ANI-2x          &       0.60\%  &               1.23\%          \\
ANI-2x RXRX 1    &       0.60\%  &               1.23\%          \\
ANI-2x RXRX 2    &       0.58\%  &               1.25\%          \\
ANI-2x RXRX 3    &       0.57\%  &               1.23\%          \\
ANI-2x RXRX 4    &       0.62\%  &               1.25\%   \\
\hline 
MACE RXRX XXS    &       X       &               X        \\
MACE RXRX XXXS   &       X       &               X        \\
\hline
\multicolumn{3}{|c|}{Trained on other datasets} \\
\hline
AIMNET2             &   0.58\%  &               1.56\%   \\
B97-3c RXRX 1        &   0.65\%  &               1.63\%   \\
B97-3c RXRX 2        &   1.08\%  &               1.72\%   \\
MACE-OFF23 L        &   0.59\%  &               1.56\%   \\
MACE-OFF23 M        &   0.54\%  &               1.55\%   \\
MACE-OFF23 S        &   X       &               X        \\
\hline
\end{tabular}

\end{table}

Molecular dynamics (MD) simulations are a common tool in the computational drug discovery processes to assess the structure and dynamics of compounds and protein-ligand complexes. MD simulations generate conformations according to their Boltzmann probability by calculating the forces and integrating Newton's equation over time. Therefore, stable MD simulations require that there are no abrupt changes of the forces, which might be caused by discontinuities in the potential energy function of the NNP. Thus, MD simulations are a simple quality check of both energies and forces after model training, as recent studies have shown that low errors of gas phase energies and forces are sometimes insufficient for model stability.\cite{stocker2022robust}

The stability benchmark includes an MD simulation of the benchmark molecule shown in Fig.~\ref{fig:gastest}, which is a mixture of structural elements of six different drugs and one peptide. The artificial benchmark molecule was created to confront the models with a molecule that is not part of any training set. An elevated temperature of $400$\,K was used to see whether the molecule falls apart. To check for large deviations of the molecular structure, the values of all covalent bond lengths ($r$) and bond angles ($\theta$) were tracked during the simulation and compared to corresponding equilibrium values of the GAFF2 force field.\cite{gaff2} Small deviations from the equilibrium values are expected due to the increased temperature and because of the different reference QM calculations. Because of the presence of S, F, and Cl atoms in the benchmark molecule, the ANI-1ccx could not be employed. 

As shown in Table~\ref{tab:gastest}, most tested machine learning potentials lead to stable MD simulations. As can be seen in the first column of Table~\ref{tab:gastest}, the average deviations of the average bond length are small (about $0.6\%$), except for the B97-3c RXRX 2 model which shows a deviation of $1.08$\%. The model training of B97-3c RXRX 2 was stopped earlier than the other model (B97-3c RXRX 1). Therefore, the slightly increased deviation might an effect of the early stopping criterion. When looking at the differences of the bond angles in the second column of Table~\ref{tab:gastest}, the models show similar deviations (between $1.2$ and $1.7$\%). 

Only three of the tested models failed to produce stable trajectories: MACE RXRX XXXS, which become unstable after 10\,ps, MACE RXRX XXS, which became unstable within 1\,ps, and the published MACE-OFF23 S model, which became unstable after 44\,ps. All three  models share the MACE architecture and use a reduced number of parameters. However, the MACE-OFF23 M and L models pass the tests, which indicates that the reduced number of parameters in the S, XXS, and XXXS models might cause the instabilities. The results suggest that it is better to use either the ANI and AIMNet2 architectures, or the larger versions of the MACE model (M or L). 

\subsubsection{Water dimer distance scans}

\begin{figure}[ht!]
\caption{Potential energy scans of the O-O distance (r) of a water dimer in the gas phase. The scans aim at testing the van der Waals repulsion (logarithmic plots) and hydrogen bond strength (insets). Methods with inadequate van der Waals repulsion might not be suitable for drug discovery methods that test for steric clashes (like molecular docking). Two different quantum-mechanical reference results were used. The first quantum-mechanical method is $\omega$B97X/6-31G(d), which is the level of theory that was used to generate the ANI-2x dataset. The second quantum-mechanical method is B97-3c/def2-mTZVP, which is the level of theory that was used to generate the ANI-2x/B97-3c dataset. a) Global energy minimum at the $\omega$B97X/6-31G(d) level of theory with an optimal r of  0.285\,nm. b) Potential energy scan of published models. The global energy minimum of the MACE models resides at a distance below 0.02\,nm c) Potential energy scan of in-house models that were trained on the ANI-2x/$\omega$B97X dataset. The global energy minima of the MACE RXRX XXS and XXXS models are below 0.12\,nm. d) Potential energy scan of in-house models that were trained on the ANI-2x/B97-3c  dataset. 
 }\label{fig:dimer}
    \begin{tabular}{cc}
    a) & b) \\ 
       \includegraphics[width=0.45\linewidth]{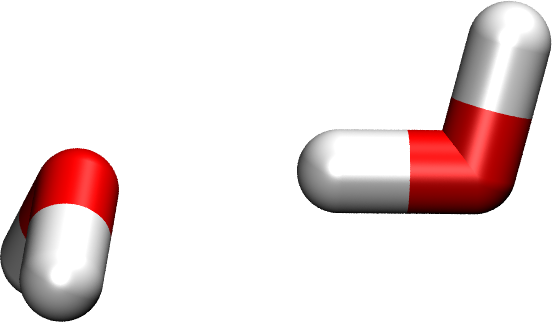} &
       \includegraphics[width=0.45\linewidth]{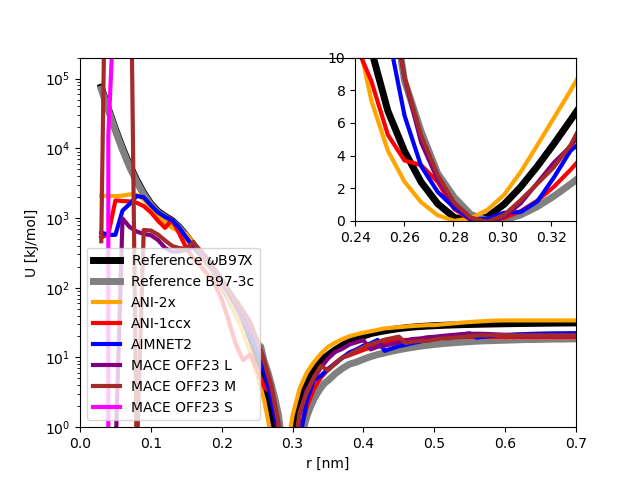} \\ 
    c) & d) \\ 
      \includegraphics[width=0.45\linewidth]{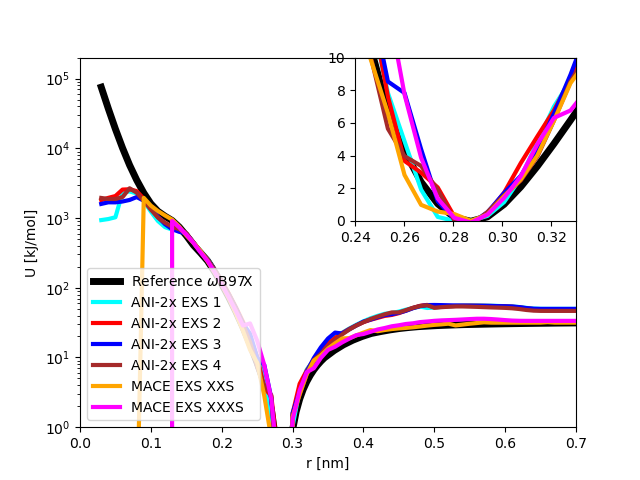}   &
      \includegraphics[width=0.45\linewidth]{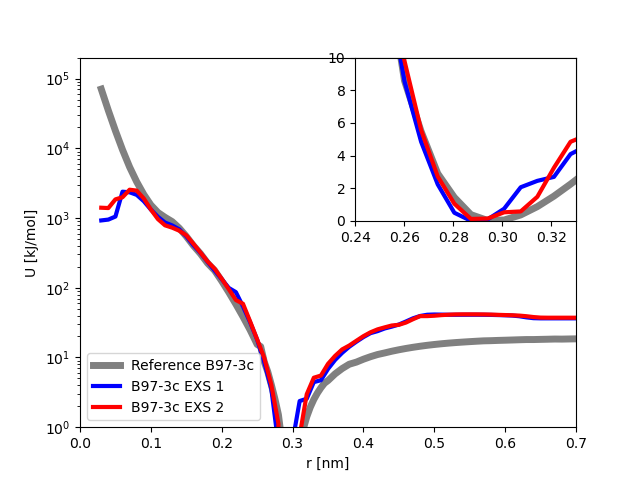}   
   \end{tabular}
\end{figure}

One of the most important intermolecular interaction types in computational drug discovery are hydrogen bonds. Hydrogen bonds exhibit a high directionality, which makes them very sensitive to misrepresentations in the model. Another important factor for applications like molecular docking is the correct representation of steric clashes, in order to assess the correct placement of a ligand in the binding pocket. As a simple test case for both aspects, we employed a potential energy scan of the oxygen-oxygen distance of a water dimer (see the starting structure in Fig.~\ref{fig:dimer}a). 

The reference potential energy curve was generated with quantum-mechanical calculations at the $\omega$B97X/6-31G(d) level of theory, which was also used to generate the ANI-2x dataset (black line in Fig.~\ref{fig:dimer}b-d). To show the dependence on the level of theory, a second reference potential energy curve was generated with B97-3c/def2-mTZVP (gray line in Fig.~\ref{fig:dimer}b-d), which was used to generate the B97-3c dataset that forms the basis for the two additional in-house NNPs. Both quantum-mechanical reference curves show a global energy minimum at a distance of $0.29$\,nm (see inset) and energies above $10^3$\,kJ/mol at distances below $0.1$\,nm (see logarithmic y-axis of plots). However, there the energy minimum with B97-3c is slightly shifted to longer distances and more shallow at longer distances than $\omega$B97X.     

Starting with the published NNPs, the potential energy curve of ANI-2x (orange line in Fig.~\ref{fig:dimer}b) matches the $\omega$B97X reference results relatively well. There is only a slight shift around the energy minimum (shown in the inset). The other published NNPs show a higher similarity to the B97-3c reference results (shown in gray), as their energy minima are flatter at longer distances. This can probably be explained by the use of different functionals and basis sets in their model training. Overall, the NNPs agree relatively well with the reference data. However, all three MACE-OFF23 models (purple, brown,  magenta lines) show additional energy minima at distances below $0.1$\,nm. Those energy minima are deeper than the global energy minimum of the reference results at 0.29\,nm. Thus, molecular docking or global energy optimization algorithms with those NNPs might lead to nonphysical results. In addition, special care has to be taken during system setup, as energy minimizations of initial structures with steric clashes might lead to even stronger steric clashes.       

The results of the in-house models trained on the ANI-2x dataset are shown in Fig.~\ref{fig:dimer}c. Most of the in-house models are very similar to the official ANI-2x model in Fig.~\ref{fig:dimer}b. Some of the models also exhibit little dents in the potential energy surface (e.g., the ANI-2x RXRX 3 at $0.27$\,nm). The long-range interactions of all ANI-based in-house models are consistenly too weak, as can be seen by the higher potential energies beyond $0.31$\,nm. Since the in-house models were trained on the same dataset and with the same data split, this is probably an expression of the possible random fluctuations during model training. Notably, the two in-house MACE models (orange and magenta lines) also show global energy minima at distances below $0.12$\,nm. Although the published and the in-house MACE models are based on different training sets, both lack the capability to correctly identify steric clashes. This indicates that this is an effect of the model architecture.  

To test the effect of using a different data set, the two models based on B97-3c are included in the Fig.~\ref{fig:dimer}d. As a reference, the quantum-mechanical potential energy scans with B97-3c are included in gray. Both models reproduce the reference potential energy surface relatively well for distances below $0.3$\,nm. However, the long range interactions are weaker than in the reference potential, which can be seen from the higher potential energies between $0.3$ and $0.7$\,nm. The weak long range interactions are very similar to the other in-house models in Fig.~\ref{fig:dimer}c. Since this distance range is longer than a hydrogen bond, this might be an expression of some small deficiencies of the dispersion or electrostatic interactions, or, alternatively, might be an effect of the $0.5$\,nm cutoff of the models. To evaluate those aspects, further tests were conducted in the condensed phase.    

\subsection{Condensed phase}  
\subsubsection{Simulations of water}

\begin{table}[ht!]
\caption{Density of water boxes and the associated RMSE of the O-O radial distribution function (RDF) with respect to
experiment. As a reference, the error of molecular mechanics simulations with TIP3P water are shown in the first row. The other rows show the results for the considered ML potentials.}
    \label{tab:watdens}
\centering
\begin{footnotesize}
\begin{tabular}{|l|c|c|}
\hline
\textbf{Model}	                      &	\textbf{Density} (g/ml)	&	\textbf{RMSE RDF}	\\
\hline

\multicolumn{3}{|c|}{Molecular Mechanics} \\ 
\hline
TIP3P   	                  &	1.00    &	0.12	\\
\hline
\multicolumn{3}{|c|}{Trained on ANI-2x dataset} \\ 
\hline
ANI-2x          &	1.06 &	0.31 \\
ANI-2x RXRX 1	&	1.41 &  0.21 \\  
ANI-2x RXRX 2	&	1.37 &	0.30 \\
ANI-2x RXRX 3	&	1.44 &	0.26 \\
ANI-2x RXRX 4	&	1.42 &	0.29 \\
\hline 
MACE RXRX XXS 	&	1.41 &	0.20\\
MACE RXRX XXXS	&	1.01 &	0.39 \\
\hline
\multicolumn{3}{|c|}{Trained on other datasets} \\ 
\hline
AIMNET2	        &  0.74  &	0.20 \\
ANI-1ccx 	&  1.13  &	0.42 \\
B97-3c RXRX 1	&  1.15  &	0.16 \\
B97-3c RXRX 2	&  0.93  &	0.08 \\
MACE-OFF23 L	&  1.09  &	0.17 \\
MACE-OFF23 M	&  1.17  &	0.17 \\
MACE-OFF23 S	&  1.11  &	0.12 \\

\hline
\end{tabular}
\end{footnotesize} \\
\end{table}

Water is the most important molecule in computational drug discovery, as the vast majority of protein-ligand binding events happen in aqueous solution. Water-water interactions are the driving force for the hydrophobic effect, which makes protein-ligand binding very sensitive to misrepresentations of the aqueous environment. For example, too strong water-water interactions in the model artificially increase the surface tension, which then might force the ligand out of aqueous solution and into hydrophobic areas like the a binding pocket. Calculating the density of water tests whether the NNPs correctly reproduce the strength of the hydrogen bonds between water molecules, as well as the balance with van der Waals interactions. The densities of water at $300$\,K and $1$\,bar with the tested NNPs  are listed in the first column of Table~\ref{tab:watdens}. The experimental reference value $1.00$\,g/ml, which is also reproduced by molecular mechanics calculations using the TIP3P water model. Among the published NNPs, ANI-2x exhibits the smallest deviation from the experimental density with 6\%, followed by the MACE-OFF23 L model with 9\%. The largest discrepancy among the published models is observed for AIMNET2, with a deviation of -$26$\%. The too low density of AIMNET2 are an indicator that the water molecules are more spread out, either because of a too large van der Waals radius, a too low hydrogen bond strength, or incorrect long-range interactions.    

The in-house models that were trained on the ANI-2x dataset exhibit deviations between 37\% and 44\% from the experimental density of water. Those deviations are noticeably higher than the ones observed for the published models. The too high densities might be caused by a too low van der Waals radius, too strong hydrogen bonds, or incorrect long range interactions. One notable exception is the MACE RXRX XXXS model, which exhibits the lowest deviation from experiment among all tested models, with an error of just 1\%. The two in-house models from the B97-3c dataset also exhibit only errors between -7\% and 15\%. This indicates that the large deviations of the density of the ANI-2x RXRX models and the MACE RXRX XXS model are not an artifact of the training procedure itself.

\begin{figure}[ht!]
\caption{Radial distribution function (RDF) of the O-O distance (r) of water in the condensed phase at 300K. The scans aim at testing the intermolecular interactions and hydrogen bond formation. The reference results were determined experimentally. a) RDFs from experiment (black), the published ANI-2x model (red), and molecular mechanics (MM) with the TIP3P water model (blue).  b) RDFs of published models based on other training sets. c) RDFs of in-house models that were trained on the ANI-2x/$\omega$B97X dataset.  d) RDFs of in-house models that were trained on the ANI-2x/B97-3c  dataset. 
 }\label{fig:rdf}
    \begin{tabular}{cc}
    a) & b) \\ 
       \includegraphics[width=0.45\linewidth]{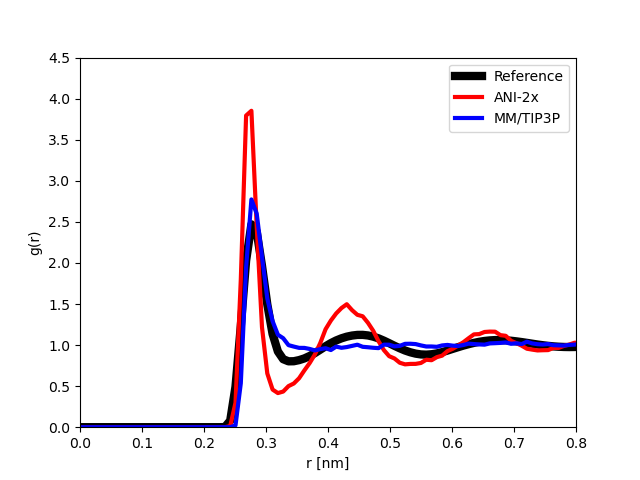} &
       \includegraphics[width=0.45\linewidth]{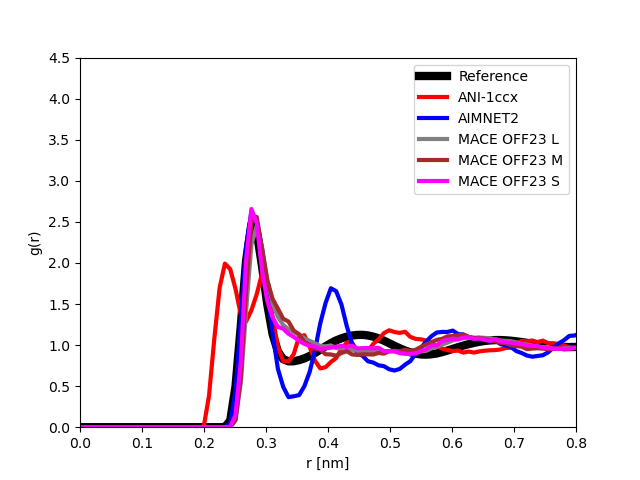} \\ 
    c) & d) \\ 
      \includegraphics[width=0.45\linewidth]{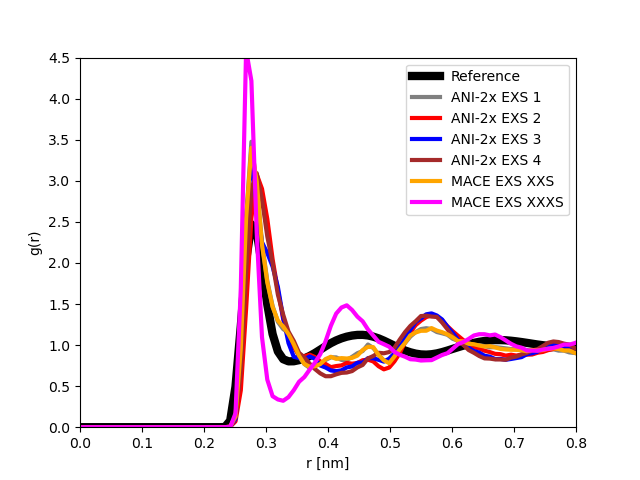}   &
      \includegraphics[width=0.45\linewidth]{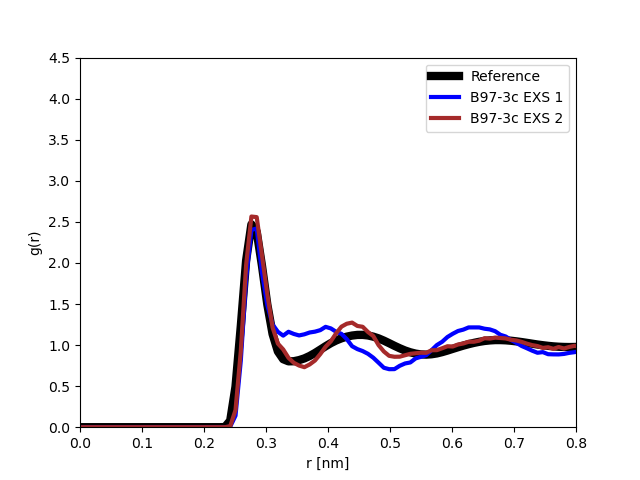}   
   \end{tabular}
\end{figure}

To further investigate the cause of the observed deviations of the density, the radial distribution functions (RDF) of the oxygen-oxygen distances between water molecules were calculated (see Fig.~\ref{fig:rdf}). The RDFs show the probability of finding two water oxygens at a certain distance. The experimental reference results are shown in black. Short distances are disallowed because of Pauli repulsion. The main peak at around 0.28\,nm corresponds to the first hydration shell. The magnitude of the peak is an indicator for the hydrogen bond strength, where higher peaks indicate stronger hydrogen bonds. The second and third hydration shells are located at about 0.45 and 0.67\,nm. Their peaks are an indicator for the interaction strength with the next-next neighbor and the next-next-next neighbor, which is a measure for the long-range interactions and the effect of polarization.  

As a comparison, the RDF of TIP3P water based on molecular mechanics is shown in blue in Fig.~\ref{fig:rdf}a. TIP3P water reproduces the first hydration shell very well, but otherwise leads to a homogenous distribution, which is characterized by the relatively flat RDF. No second or third hydration shell is noticeable. Calculating the root mean square deviation with respect to the experimental RDF leads to 0.12 (see second column of Table~\ref{tab:watdens}). Considering the increased computational costs of the NNPs, one would expect a higher accuracy than the molecular mechanics result. 

The RDF of the published ANI-2x model is shown in red in Fig.~\ref{fig:rdf}a. Compared to the experimental reference, the peaks of the first hydration shell is significantly higher. This indicates stronger hydrogen bonds. Also the second hydration shell exhibits a higher peak than the experimental reference, while the minimum between the peaks is much lower. This is an indicator that ANI-2x water is more structured and solid than liquid water. A visual inspection of the simulation indicates that ANI-2x water is in an amorphous solid state. 

\begin{figure}
    \centering
    \includegraphics[width=0.8\linewidth]{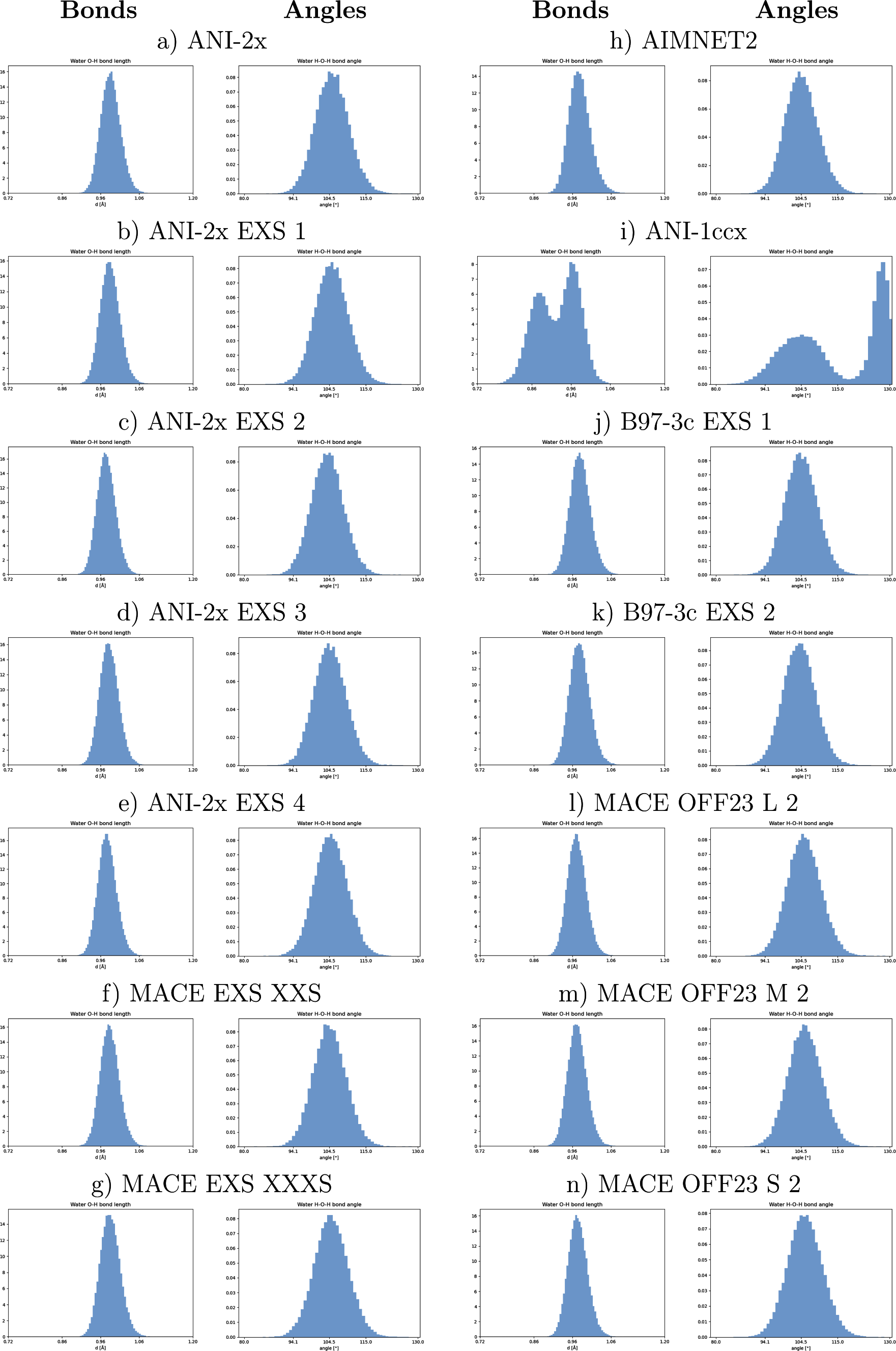}
    \caption{Distributions of the O-H bond lengths and H-O-H bond angles of water throughout the MD simulation. The plotting range of the bond lengths is between $0.072$ and $0.12$\,nm, while the bond angles are displayed between 78.4 and 130.6\degrees{}. In the presence of other water molecules, ANI-1ccx (subplot i) displays nonphysical extra peaks, which lead to a distorted water structure.}
    \label{fig:waterbondsangles}
\end{figure}

The results for other published models are shown in Fig.~\ref{fig:rdf}b. The RDF of ANI-1ccx (red line) exhibits a split first peak, where a part of the water molecules are at a too close distance. This could indicate that the hydrogen bond interactions are partially able to overpower the van der Waals repulsion. However, a closer inspection of the O-H bond and H-O-H angle distributions in Fig.~\ref{fig:waterbondsangles}i reveals that, in the presence of other water molecules, ANI-1ccx exhibits additional peaks in the bond length and bond angle distribution. The O-H bond length becomes shorter, and the H-O-H angle flatter. The changes of the bonded structure distort the hydrogen bonding network.  The RDF from AIMNET2 reproduces the first hydration shell very well. However, the peak of the second hydration shell is relatively high, and the minimum between the first and second hydration shell is relatively low. This is an indicator that AIMNET2 water is not liquid at room temperature. Like ANI-2x, a visual inspection of the AIMNET2 trajectory reveals an amorphous solid state. The MACE-OFF23 models reproduce the first and third hydration shell relatively well. However, there is no peak for the second hydration shell. 

As shown in Fig.~\ref{fig:rdf}c, the in-house ANI-2x models, as well as the MACE RXRX XXS model, reproduce the first hydration shell relatively well. This is surprising, given that the official ANI-2x model based on the same dataset exhibits a high peak for the first hydration shell. In addition, the in-house models are liquid at room temperature, which is not the case for ANI-2x. However, instead of a second and third hydration shell at the experimental positions, the in-house models only exhibit a second hydration shell at around $0.56$\,nm. One notable exception among the in-house models is MACE RXRX XXXS. Although it uses the same architecture as MACE RXRX XXS, and is based on the same training set as the other ANI-2x models, it shows a very different behavior. The MACE RXRX XXXS shows the highest peak for the first hydration shell among all tested models. In addition, there is a high peak for the second hydration shell and a low minimum between the first and second hydration shell. Like the published ANI-2x model, the MACE RXRX XXXS is not liquid at room temperature and forms an amorphous solid state. Overall, the RDF of the MACE RXRX XXXS is very similar to the RDF of the published ANI-2x model. This is and indicator that different architectures can yield relatively similar results if they are trained on the same dataset.

As an additional test, also the two in-house models that were trained on the B97-3c data set are included in Fig.~\ref{fig:rdf}d. Both in-house B97-3c correctly reproduce the first hydration shell. The  B97-3c RXRX 1 model does not lead to clear second hydration shell, showing a plateau after the first hydration shell instead. However, the B97-3c RXRX 2 model perfectly reproduces all three hydration shells. The difference between B97-3c RXRX 1 and 2  radial distribution functions around $0.35$\,nm is somewhat surprising, given that the water dimer scans show good agreement between the two models at this distance (see Fig.~\ref{fig:dimer}d). This indicates that subtle differences in the hydrogen bonding strength, as well as many-body effects, can have a significant impact on the nonbonded interactions in the condensed phase. Nevertheless, the high accuracy of the B97-3c RXRX 2 model is an indicator that the employed training procedure is correct. In addition, the B97-3c RXRX 2 model is the only tested NNP that leads to a lower error than the molecular mechanics TIP3P water model, showing an RMSE of just 0.08 compared to the experimental RDF (see Table~\ref{tab:watdens}). This shows that NNPs are able to out-compete highly optimized molecular mechanics parameters.

\section{Conclusions} 
\label{sec:conclusions}

Based on the RMSE for the test set, all in-house models, except for MACE RXRX XXXS, performed very similarly, with RMSE between 1.8 and 1.9\,\kcal{}. However, the other tests in the gas phase and in the condensed phase showed large variations of model quality, which shows that the RMSE of single point energies is not a reliable metric for model quality. As a first test, a normal mode analysis of 14 benchmark molecules was used to study the quality of the energy surfaces far away from the energy minima. The RMSE of the normal mode analysis show a very different ranking of the models than the RMSE from the test set. In addition, the normal mode analysis revealed large discrepancies of the MACE OFF S model for large displacements from the energy  minimum. The molecular dynamics simulations of the drug-like benchmark molecule in the gas phase demonstrated that most ML potentials lead to stable simulations and small deviations of the equilibrium bond and angle terms. However, the two in-house models MACE RXRX XXS and XXXS, as well as the published MACE-OFF23 S model failed the test, as they lead to unstable simulations. Those models share the MACE architecture with a small parameter set. However, the larger MACE models (M and L) did pass the test, so the unstable simulations can probably be explained by the use of a smaller set of parameters. Therefore, MACE models with at least size M are recommended. One model based on the ANI architecture, B97-3c RXRX 2, also showed slightly higher deviations of the bond and angle terms (1.08 and 1.72\%, respectively), which can possibly be attributed to the shorter model training compared to the other models (1500 instead of about 2500 epochs).   

The potential energy scans of water dimers revealed that all MACE models fail when it comes to correctly representing steric clashes. Their global energy minimum resides within the van der Waals radius of oxygen (distances below 0.12\,nm). This indicates that additional care must be taken when using this architecture for applications that might involve steric clashes, like molecular docking or binding affinity predictions with free energy calculations. Notably, most molecular mechanics force fields show a similar behavior (i.e., that the electrostatic attraction is able to overpower the van der Waal repulsion at  very short distances). In molecular mechanics, this problem is overcome by turning off the electrostatic component of force fields when evaluating steric clashes. Developing an analogous solution (e.g., adding a steric repulsion potential) might solve this problem of MACE models. On the other hand, several of the in-house models with the ANI architecture failed to correctly represent the long-range interactions of water dimers, which might have a negative impact on the liquid structure of water or protein-ligand interactions.   

The molecular dynamics simulations of the condensed phase of water showed that most ML potentials failed to yield the correct density of water. Only the ANI-2x, MACE RXRX XXXS, ANI-1ccx, B97-3c RXRX 1 and 2, as well as the published MACE-OFF23 models yielded deviations below 20\% with respect to the experimental density of water. An analysis of the radial distribution function of those models revealed further discrepancies. The simulation with ANI-1ccx waters leads to a solid phase with large deviations of the bond and angle terms of water due to the appearance of nonphysical energy minima. This emergent phenomenon could not be predicted by the gas phase water dimer data, where ANI-1ccx exhibited small deviations of water bond lengths and angles, and also correctly reproduced the global energy minimum of the hydrogen bond. This indicates that the observed anomaly might be a multi-body effect, which only becomes apparent in tests of the condensed phase.  

Likewise, the water simulations with ANI-2x yielded an amorphous solid state at room temperature. In molecular simulations, this solid water will most likely represent a major hindrance for the correct sampling of conformation changes of solutes and misrepresent the hydrophobic effect. Interestingly, a very similar radial distribution function with an amorphous solid state was found for the MACE RXRX XXXS model. This shows that this is not an effect of the model architecture. In addition, the in-house ANI-2x RXRX models did not solidify, which indicates that this is also not an effect of the dataset, as the same data split was used for the ANI-2x RXRX and MACE RXRX models. Among all tested models, only the published MACE-OFF23 models and the in-house B97-3c RXRX models yielded adequate radial distribution functions of water (RMSE below 0.2). Notably, only the radial distribution function of the B97-3c RXRX 2 model showed a better agreement with experiment than the simple TIP3P water model based on molecular mechanics. This shows that the liquid structure of water is still a challenge for most ML potentials. Considering that the water-water interactions are the driving force of the hydrophobic effect in drug binding, the current results indicate the ML potentials will still have to improve for real-world applications in computational drug discovery. We hope that the presented results can help to point out some of the main challenges, to make sure that those problems can be addressed in future versions of machine learning potentials.

\begin{suppinfo}
Figures S1 to S7 depict the loss functions of the in-house models during model training. The results of a quasi-harmonic analysis of 14 simple benchmark molecules in the gas phase are included in the section ``Further quasi-harmonic analysis tests''.\cite{mybutanol,qmdrude,leds,bss,off2,Eastman2023a,Chow1995,mdtraj,bro95a} The average deviations of the bond lengths and angles, as well as the corresponding force constants, are reported in Table S1. Tables S2 to S5 provide additional RMSE data for the normal mode analysis in Table~\ref{tab:normalmode}, using different frequency bands of the normal modes. 
\end{suppinfo}

\acknowledgement
The authors would like to thank Marcus Wieder, as well as Stefan Boresch, and Johannes Karwounopoulos for their help and very insightful discussions.

\setstretch{1.0}
\bibliography{bib.bib}

\end{document}


\maketitle

\subsection{Model training}

\begin{figure}[h!]
\caption{Loss function of ANI-2x RXRX 1 model. 
 }\label{fig:loss1}
       \includegraphics[width=0.95\linewidth]{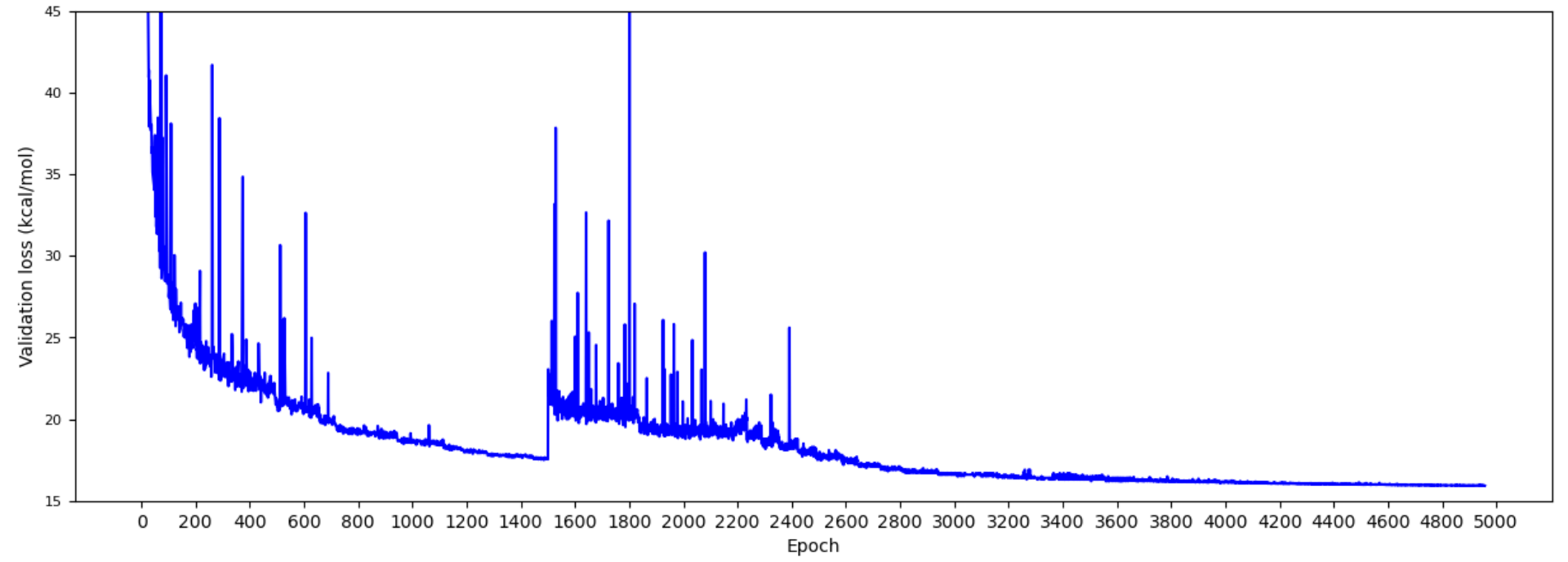} \\ 
\end{figure}

\begin{figure}[h!]
\caption{Loss function of ANI-2x RXRX 2 model. 
 }\label{fig:loss2}
       \includegraphics[width=0.95\linewidth]{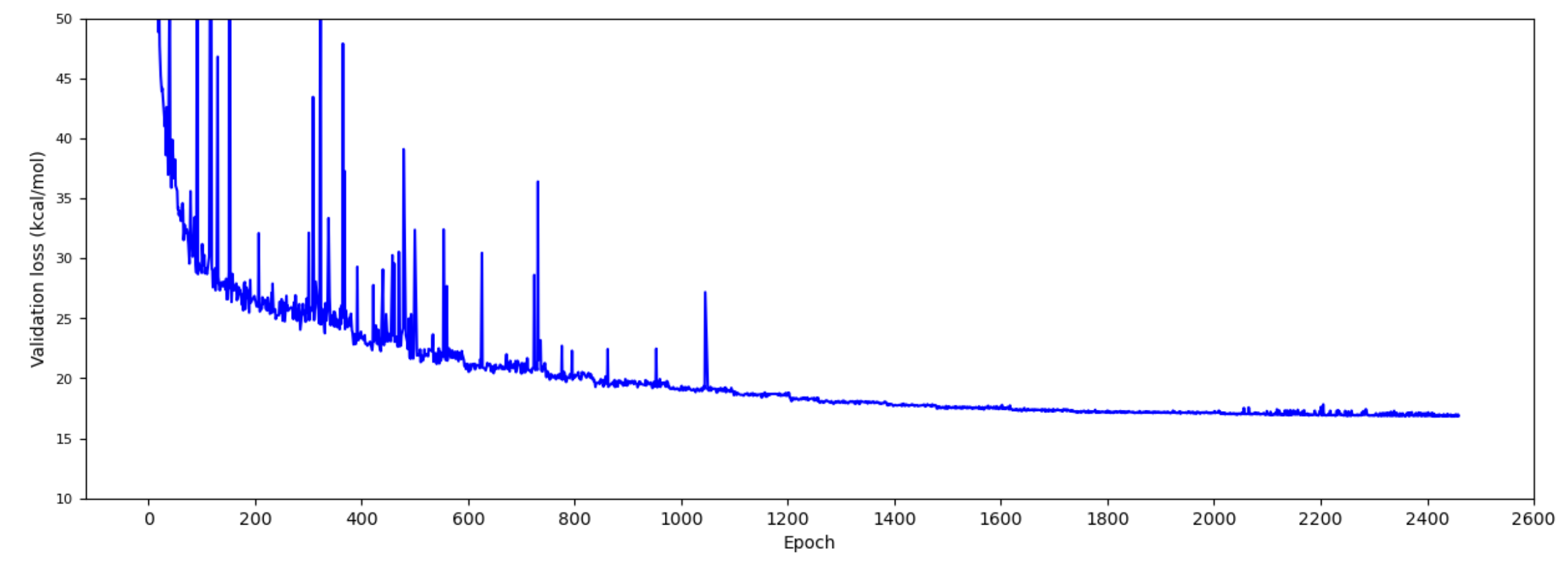} \\ 
\end{figure}

\begin{figure}[h!]
\caption{Loss function of ANI-2x RXRX 3 model. 
 }\label{fig:loss3}
       \includegraphics[width=0.95\linewidth]{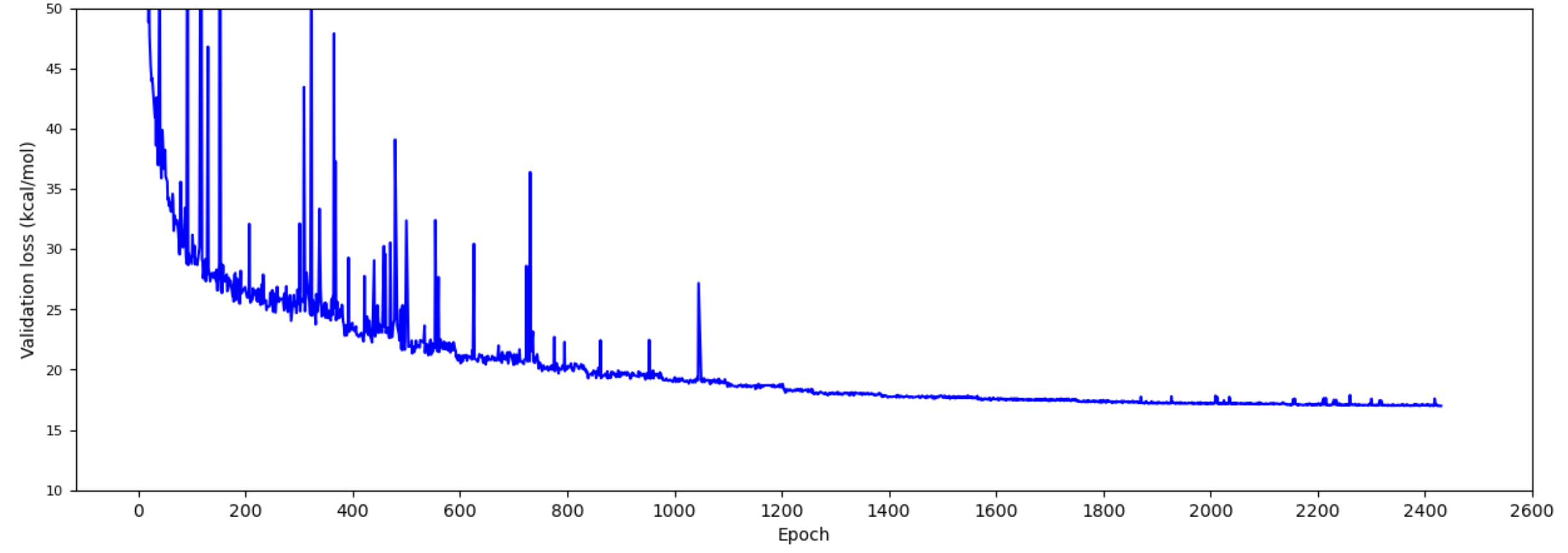} \\ 
\end{figure}

\begin{figure}[h!]
\caption{Loss function of ANI-2x RXRX 4 model. 
 }\label{fig:loss4}
       \includegraphics[width=0.95\linewidth]{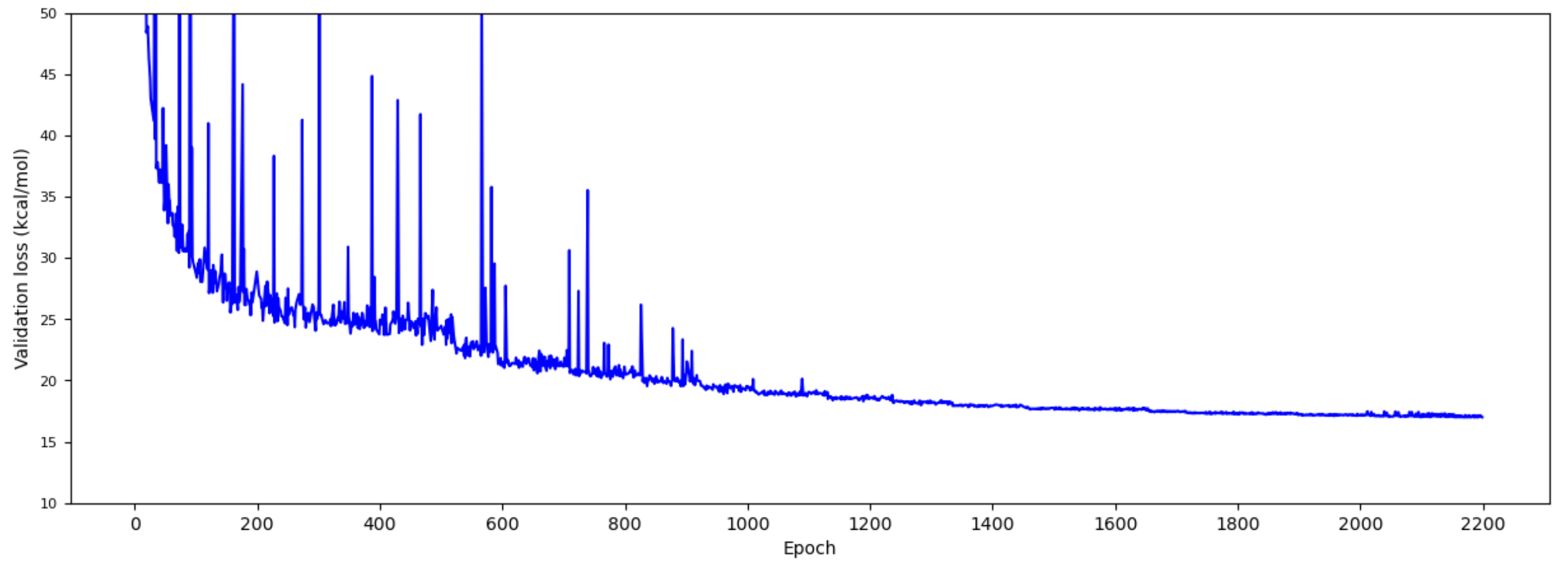} \\ 
\end{figure}

\begin{figure}[h!]
\caption{Loss function of B97-3c RXRX 1 model. 
 }\label{fig:loss5}
       \includegraphics[width=0.95\linewidth]{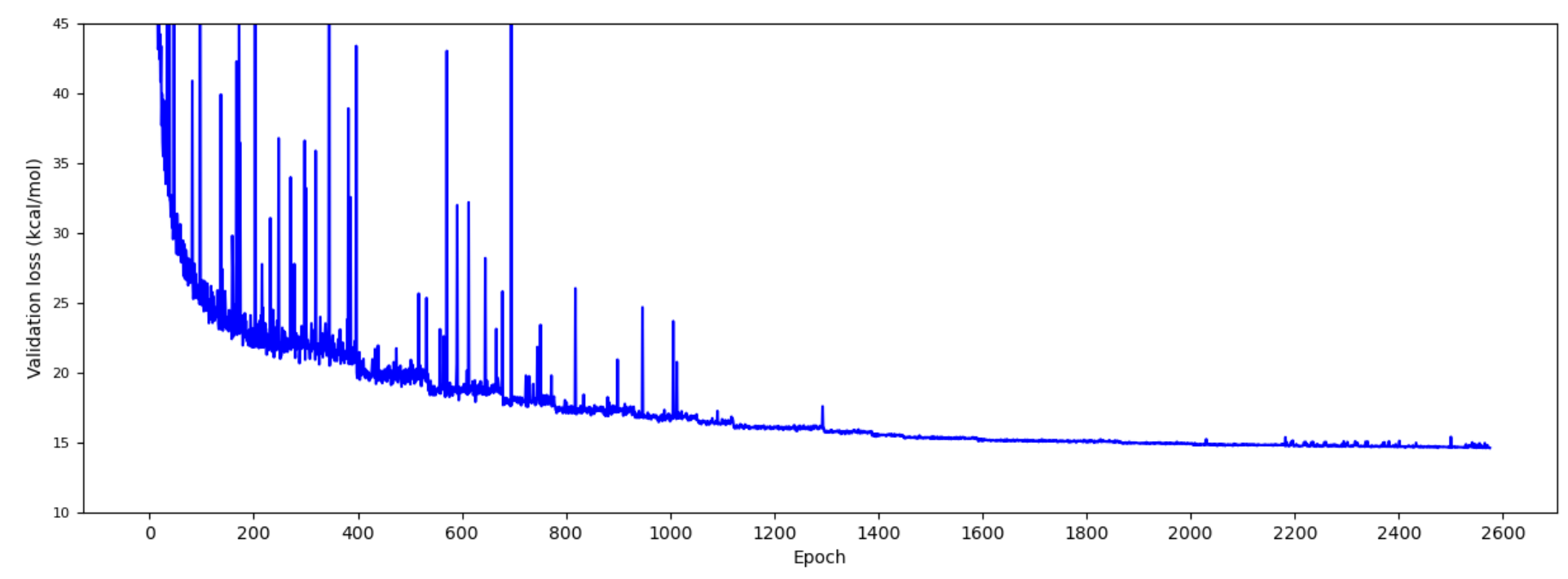} \\ 
\end{figure}

\begin{figure}[h!]
\caption{Loss function of MACE RXRX XXS model. 
 }\label{fig:loss6}
       \includegraphics[width=0.95\linewidth]{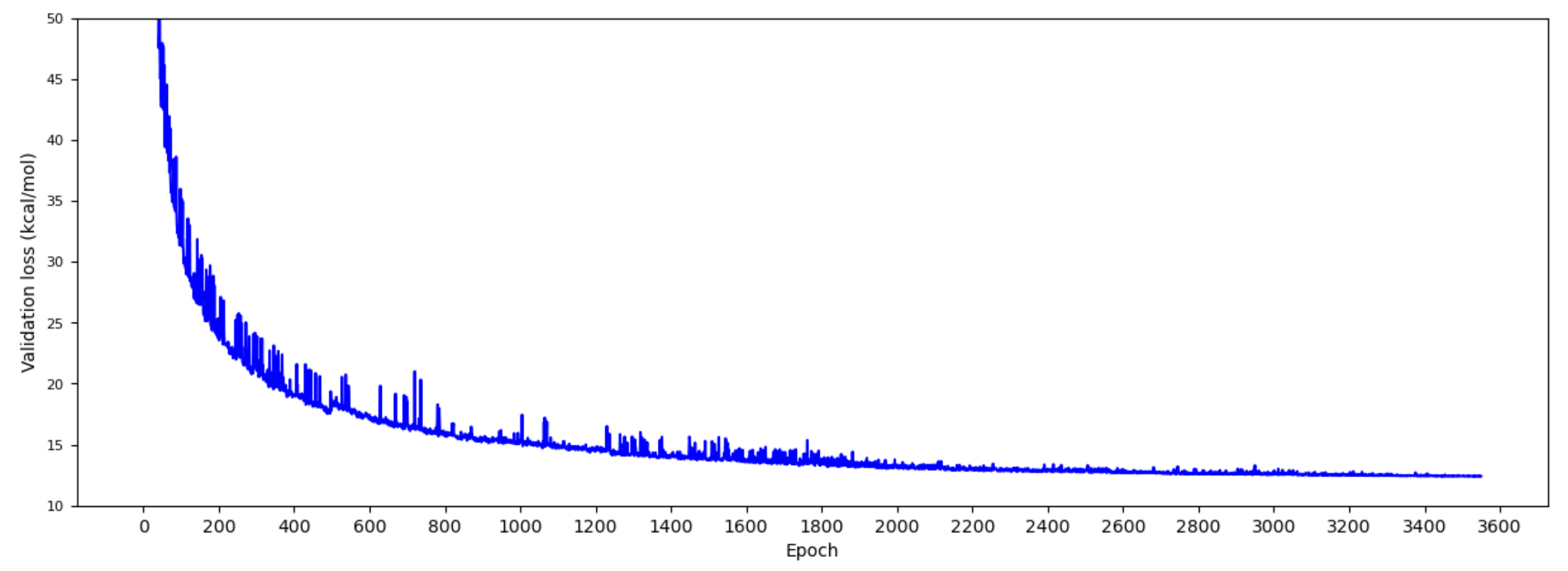} \\ 
\end{figure}

\begin{figure}[h!]
\caption{Loss function of MACE RXRX XXXS model. 
 }\label{fig:loss7}
       \includegraphics[width=0.95\linewidth]{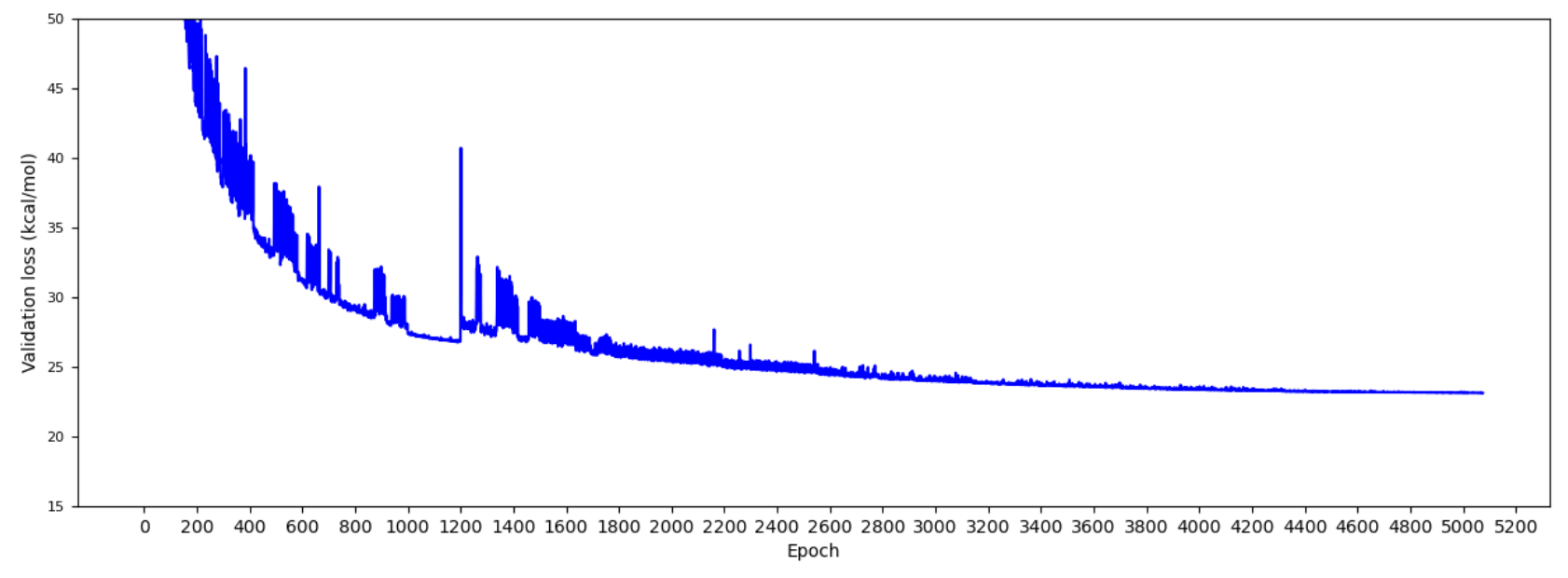} \\ 
\end{figure}

\clearpage

\subsection{Further quasi-harmonic analysis tests}

The 14 benchmark systems include water, ethane, methanol, methanethiol, ethanol, acetamide, tetrahydrofuran, n-hexane, cyclohexane, benzene, phenol, aniline, N-acetyl-alanine-methylamide, and N-acetyl-serine-methylamide (see Fig.~1 of main manuscript), and were selected from previous publications. They represent relevant chemical groups and levels of hydrophobicity and aromaticity.\cite{mybutanol,qmdrude,leds2} The benchmark molecules were generated with BioSimSpace\cite{bss} using Open Force Field 2.2.0.\cite{off2}  The centers of mass of the molecules were assigned random Cartesian coordinates between 0 and 100\,{\AA}. Before each ML simulation, the geometry of the molecules was optimized using the L-BFGS minimizer. The molecular dynamics simulations at the ML level were performed with OpenMM 8.1\cite{Eastman2023a} at an elevated temperature of $400$\,K to test the stability of the molecules. Langevin dynamics were performed with a friction constant of $1$\,ps$^{-1}$.\cite{Chow1995} The time step was $1.0$\,fs and the simulation length $0.25$\,ns. The trajectory was saved every $0.5$\,ps. The distributions of all bond length and bond angle values during the simulation were calculated using MDTraj.\cite{mdtraj} The quasi-harmonic analysis of the individual distributions was performed as outlined in Ref.~\citenum{bro95a}.

\begin{table}[]
\caption{Quasiharmonic analysis of molecular dynamics simulations of the 14 benchmark molecules in the gas phase. All models lead to stable molecular dynamics simulations. The average relative deviations ($\langle \Delta \rangle$) over all chemical bonds and angles with respect to the official ANI-2x model are given in percent. Both the equilibrium values of the bond length ($r_{eq}$) and bond angle ($\theta_{eq}$), as well as the corresponding force constants (K) are listed.}\label{tab:gastest}
\begin{tabular}{|l|l|l||l|l|}
\hline
\textbf{Model}     & \textbf{$\mathbf{\langle \Delta r_{eq} \rangle}$}  & \textbf{$\mathbf{\langle \Delta K_{r} \rangle}$} & \textbf{$\mathbf{\langle \Delta \theta_{eq} \rangle}$}   & \textbf{$\mathbf{\langle \Delta K_{\theta} \rangle}$}  \\
\hline  
\multicolumn{5}{|c|}{Trained on ANI-2x dataset} \\ 
\hline 
ANI-2x RXRX 1	&	0.16\%	&	7.41\%	&	0.23\%	&	7.56\%	\\
ANI-2x RXRX 2	&	0.14\%	&	7.38\%	&	0.25\%	&	7.93\%	\\
ANI-2x RXRX 3	&	0.15\%	&	8.16\%	&	0.23\%	&	7.00\%	\\
ANI-2x RXRX 4	&	0.16\%	&	8.15\%	&	0.25\%	&	6.72\%	\\
MACE RXRX XXS 	&	0.15\%	&	8.06\%	&	0.25\%	&	7.39\%	\\
MACE RXRX XXXS	&	0.17\%	&	8.98\%	&	0.29\%	&	7.72\%	\\
\hline 
\multicolumn{5}{|c|}{Trained on other datasets} \\ 
\hline 
AIMNET2	            &	0.46\%	&	9.30\%	&	0.30\%	&	7.76\%	\\
ANI-1ccx$^\#$	    &	0.31\%	&	9.12\%	&	0.31\%	&	8.11\%	\\
B97-3c RXRX 1	    &	0.39\%	&	9.25\%	&	0.29\%	&	7.47\%	\\
B97-3c RXRX 2	    &	0.39\%	&	9.63\%	&	0.29\%	&	8.23\%	\\
MACE OFF23 L	    &	0.47\%	&	10.81\%	&	0.29\%	&	7.91\%	\\
MACE OFF23 M	    &	0.47\%	&	8.52\%	&	0.29\%	&	7.91\%	\\
MACE OFF23 S	    &	0.49\%	&	9.04\%	&	0.27\%	&	8.50\%	\\
\hline
\end{tabular}
\flushleft $^\#$ Results without methanethiol because of missing parameters for sulfur   
\end{table}

Molecular dynamics (MD) simulations are a common tool in the computational drug discovery processes to assess the structure and dynamics of compounds and protein-ligand complexes. MD simulations generate conformations according to their Boltzmann probability by calculating the forces and integrating Newton's equation over time. Therefore, stable MD simulations require that there are no abrupt changes of the forces, which might be caused by discontinuities in the potential energy function of the NNP. Thus, MD simulations are a simple additional quality check of both energies and forces after model training. 

The stability benchmark includes an MD simulation of the 14 small molecules in Fig.~1 of the main manuscript at an elevated temperature of $400$\,K. To check for large deviations of the molecular structure, the values of all covalent bond lengths ($r$) and bond angles ($\theta$) were tracked during the simulation and compared to the average results of MD simulations with ANI-2x under the same conditions. The corresponding force constants ($K$) were calculated from the variance of the 167 bond length and 280 bond angle values. The average relative deviations of all average bond lengths, {${\langle \Delta r_{eq} \rangle}$}, all bond force constants, {${\langle \Delta K_{r} \rangle}$},  all average bond angles,  {$\ {\langle \Delta \theta_{eq} \rangle}$},  and all angle force constants,  {${\langle \Delta K_{\theta} \rangle}$, are shown in Table~\ref{tab:gastest}. 

All tested machine learning potentials lead to stable MD simulations. As can be seen in the first and third column of Table~\ref{tab:gastest}, the average deviations of the average bond lengths and bond angles are negligible ($<0.5\%$). The average deviations of the average force constants in the second and fourth column of Table~\ref{tab:gastest} are slightly higher with up to $10.8\%$, but this can probably be explained by the higher uncertainty of the underlying variance values and the fluctuations in the temperature control. Overall, slightly higher deviations are observed for the NNPs that are not trained on the ANI-2x dataset, which can be explained by the use of a different quantum-mechanical level of theory during model training. The good performance of all NNPs is reassuring, considering that low errors of gas phase energies and forces at the end of model training are sometimes insufficient for model stability.\cite{stocker2022robust}

\subsection{Normal mode analysis results with different cutoffs}

\begin{table}[!ht]
    \centering
    \caption{Root mean square errors (RMSE) from normal mode analysis of the 14 benchmark molecules using a cutoff of 50\,\kcal{}. The columns list the model, the level of theory for the reference calculations, the RMSE for all wave numbers $\nu$, the RMSE for $\nu>1800$, $1800 < \nu >=  1000$, $1000 < \nu >=  500$, and $500 < \nu$.  }\label{tab:nm50}
    \begin{tabular}{|lllcccc|}
    \hline
        Model & Reference & All $\nu$ & $>1800$ &	$1800-1000$	& $1000-500$  & $<500$   \\	
    \hline 	
        ANI-2x & $\omega$B97X/6-31G(d) & 0.462 & 0.121 & 0.495 & 0.744 & 0.510  \\ 
        ANI-2x Model 1 & $\omega$B97X/6-31G(d) & 0.538 & 0.162 & 0.566 & 0.827 & 0.655  \\ 
        ANI-2x Model 2 & $\omega$B97X/6-31G(d) & 0.561 & 0.184 & 0.603 & 0.793 & 0.707  \\ 
        ANI-2x Model 3 & $\omega$B97X/6-31G(d) & 0.548 & 0.169 & 0.585 & 0.841 & 0.640  \\ 
        ANI-2x Model 4 & $\omega$B97X/6-31G(d) & 0.656 & 0.161 & 0.711 & 1.057 & 0.703  \\ 
        ANI-2x B97-3c & B97-3c/def2-TZ & 0.713 & 0.367 & 0.740 & 1.121 & 0.719  \\ 
        ANI-2x B97-3c & B97-3c/def2-TZ & 0.693 & 0.352 & 0.853 & 0.948 & 0.660  \\ 
        MACE RXRX XXXS & $\omega$B97X/6-31G(d) & 0.436 & 0.119 & 0.415 & 0.740 & 0.559  \\ 
        MACE RXRX XXS & $\omega$B97X/6-31G(d) & 0.676 & 0.167 & 0.677 & 1.041 & 0.911  \\ 
        MACE OFF S & wB97/TZ & 1.199 & 0.682 & 1.359 & 1.462 & 1.374  \\ 
        MACE OFF M & wB97/TZ & 0.987 & 0.633 & 0.764 & 1.728 & 1.159  \\ 
        AIMNET2 & B97-3c/def2-TZ & 0.883 & 0.511 & 0.880 & 1.209 & 1.121 \\ \hline
    \end{tabular}
\end{table}

\begin{table}[!ht]
    \centering
    \caption{Root mean square errors from normal mode analysis of the 14 benchmark molecules using a cutoff of 100. The columns list the model, the level of theory for the reference calculations, the RMSE for all wave numbers $\nu$, as well as the RMSE for $\nu$>1800\,\invcm{}, 1800\,\invcm{} $< \nu >=$  1000\,\invcm, 1000\,\invcm $< \nu >=$   500\,\invcm, and 500\,\invcm $< \nu$. }\label{tab:nm100}
        \begin{tabular}{|lllcccc|}
    \hline
        Model & Reference & All $\nu$ & $>1800$ &	$1800-1000$	& $1000-500$  & $<500$   \\	
    \hline     
        ANI-2x & $\omega$B97X/6-31G(d) & 0.997 & 0.274 & 0.834 & 2.073 & 0.783  \\ 
        ANI-2x Model 1 & $\omega$B97X/6-31G(d) & 1.121 & 0.271 & 0.940 & 2.286 & 0.992  \\ 
        ANI-2x Model 2 & $\omega$B97X/6-31G(d) & 1.159 & 0.325 & 1.040 & 2.246 & 1.031  \\ 
        ANI-2x Model 3 & $\omega$B97X/6-31G(d) & 1.134 & 0.267 & 1.014 & 2.220 & 1.002  \\ 
        ANI-2x Model 4 & $\omega$B97X/6-31G(d) & 1.429 & 0.249 & 1.098 & 3.143 & 1.074  \\ 
        ANI-2x B97-3c & B97-3c/def2-TZ & 1.272 & 0.419 & 1.198 & 2.158 & 1.282  \\ 
        ANI-2x B97-3c & B97-3c/def2-TZ & 1.177 & 0.465 & 1.326 & 1.935 & 1.087  \\ 
        MACE RXRX XXXS & $\omega$B97X/6-31G(d) & 0.920 & 0.210 & 0.678 & 1.908 & 0.916  \\ 
        MACE RXRX XXS & $\omega$B97X/6-31G(d) & 1.210 & 0.226 & 1.086 & 2.038 & 1.544  \\ 
        MACE OFF S & wB97/TZ & 2.588 & 1.570 & 1.800 & 3.525 & 4.022  \\ 
        MACE OFF M & wB97/TZ & 1.960 & 1.048 & 1.296 & 2.862 & 2.975  \\ 
        AIMNET2 & B97-3c/def2-TZ & 1.257 & 0.584 & 1.192 & 1.918 & 1.469 \\ \hline
    \end{tabular}
\end{table}

\begin{table}[!ht]
    \centering
    \caption{Root mean square errors from normal mode analysis of the 14 benchmark molecules using a cutoff of 200. The columns list the model, the level of theory for the reference calculations, the RMSE for all wave numbers $\nu$, as well as the RMSE for $\nu$>1800\,\invcm{}, 1800\,\invcm{} $< \nu >=$  1000\,\invcm, 1000\,\invcm $< \nu >=$   500\,\invcm, and 500\,\invcm $< \nu$.  }\label{tab:nm200}    
        \begin{tabular}{|lllcccc|}
    \hline        Model & Reference & All $\nu$ & $>1800$ &	$1800-1000$	& $1000-500$  & $<500$   \\	
    \hline 
        ANI-2x & $\omega$B97X/6-31G(d) & 1.878 & 1.175 & 1.199 & 3.862 & 1.208  \\ 
        ANI-2x Model 1 & $\omega$B97X/6-31G(d) & 1.950 & 1.084 & 1.443 & 3.829 & 1.419  \\ 
        ANI-2x Model 2 & $\omega$B97X/6-31G(d) & 2.001 & 1.048 & 1.440 & 3.976 & 1.523  \\ 
        ANI-2x Model 3 & $\omega$B97X/6-31G(d) & 1.948 & 1.024 & 1.489 & 3.786 & 1.456  \\ 
        ANI-2x Model 4 & $\omega$B97X/6-31G(d) & 2.149 & 1.100 & 1.395 & 4.455 & 1.558  \\ 
        ANI-2x B97-3c & B97-3c/def2-TZ & 2.058 & 0.446 & 1.532 & 3.890 & 1.793  \\ 
        ANI-2x B97-3c & B97-3c/def2-TZ & 1.961 & 0.553 & 1.597 & 4.116 & 1.647  \\ 
        MACE RXRX XXXS & $\omega$B97X/6-31G(d) & 1.493 & 0.442 & 0.902 & 3.065 & 1.399  \\ 
        MACE RXRX XXS & $\omega$B97X/6-31G(d) & 1.860 & 0.744 & 1.335 & 3.324 & 2.240  \\ 
        MACE OFF S & wB97/TZ & 4.334 & 2.641 & 2.355 & 6.317 & 6.466  \\ 
        MACE OFF M & wB97/TZ & 4.205 & 1.223 & 1.880 & 5.238 & 7.697  \\ 
        AIMNET2 & B97-3c/def2-TZ & 1.801 & 0.613 & 1.542 & 2.882 & 2.159 \\ \hline
    \end{tabular}
\end{table}

\begin{table}[!ht]
    \centering
    \caption{Root mean square errors from normal mode analysis of the 14 benchmark molecules using a cutoff of 400. The columns list the model, the level of theory for the reference calculations, the RMSE for all wave numbers $\nu$, as well as the RMSE for $\nu$>1800\,\invcm{}, 1800\,\invcm{} $< \nu >=$  1000\,\invcm, 1000\,\invcm $< \nu >=$   500\,\invcm, and 500\,\invcm $< \nu$. }\label{tab:nm400}    
        \begin{tabular}{|lllcccc|}
    \hline        Model & Reference & All $\nu$ & $>1800$ &	$1800-1000$	& $1000-500$  & $<500$   \\	
    \hline 
        ANI-2x & $\omega$B97X/6-31G(d) & 3.301 & 1.518 & 1.860 & 7.316 & 1.240  \\ 
        ANI-2x Model 1 & $\omega$B97X/6-31G(d) & 2.775 & 1.423 & 1.860 & 5.768 & 1.520  \\ 
        ANI-2x Model 2 & $\omega$B97X/6-31G(d) & 3.247 & 1.365 & 2.503 & 6.578 & 1.603  \\ 
        ANI-2x Model 3 & $\omega$B97X/6-31G(d) & 3.289 & 1.291 & 2.583 & 6.658 & 1.545  \\ 
        ANI-2x Model 4 & $\omega$B97X/6-31G(d) & 3.120 & 1.449 & 1.749 & 6.846 & 1.573  \\ 
        ANI-2x B97-3c & B97-3c/def2-TZ & 3.744 & 0.446 & 4.284 & 5.754 & 1.811  \\ 
        ANI-2x B97-3c & B97-3c/def2-TZ & 3.024 & 0.553 & 2.323 & 6.743 & 1.981  \\ 
        MACE RXRX XXXS & $\omega$B97X/6-31G(d) & 1.973 & 0.489 & 1.187 & 4.141 & 1.510  \\ 
        MACE RXRX XXS & $\omega$B97X/6-31G(d) & 2.372 & 0.907 & 1.597 & 4.584 & 2.309  \\ 
        MACE OFF S & wB97/TZ & 75.64 & 2.641 & 4.846 & 13.70 & 185.4  \\ 
        MACE OFF M & wB97/TZ & 5.558 & 1.223 & 4.091 & 8.394 & 7.868  \\ 
        AIMNET2 & B97-3c/def2-TZ & 2.556 & 0.613 & 2.128 & 4.274 & 2.929 \\ \hline
    \end{tabular}
\end{table}

\clearpage

\subsection{SMILES of drug-like benchmark molecule}

\seqsplit{C([C@@H]1C[C@H]2[C@@H]3CCC4=C(C(=O)C=C[C@@]4([C@]3([C@H](C[C@@]2([C@]1(C(=O)CO)O)C)O)F)CCCc1nn(c2c1nc([nH]c2=O)c1c(ccc(c1)S(=O)(=O)N1CCN(CC1)CCNC(=O)[C@@H](NC(=O)C)Cc1c[nH]c2c1cccc2)O[C@@]1(C[C@H](O[C@@H]2[C@H](C(=O)O[C@H](CC)[C@@]([C@@H]([C@H](C(=O)[C@@H](C[C@@]([C@@H]([C@H]2C)O[C@@H]2O[C@H](C)C[C@@H]([C@H]2O)N(C)C)(C)OC)C)C)O)(C)O)C)O[C@H]([C@@H]1O)C)C)CCN1[C@H]2[C@@H]3C=C[C@H](O)[C@H]4[C@@]3(c3c(O4)c(ccc3C2)O)CC1)CC(=O)N[C@H]1[C@@H]2N([C@H](C(C)(S2)C)C(=O)O)C1=O)CC1=NCC(=O)N(C)c2c1cc(cc2)Cl} 

\setstretch{1.0}

\bibliography{bib.bib}